\documentclass[11pt]{article}
\usepackage{epstopdf}
\usepackage[a4paper,hmarginratio=1:1,vmarginratio=2:3,totalwidth=15.4cm,totalheight=22.75cm]{geometry}
\usepackage{bm,epstopdf,epsfig,amsmath,amssymb,amsfonts,colordvi,latexsym,wrapfig,comment,cancel,verbatim,slashed}
\usepackage{epsfig,bbm}
\usepackage{graphicx,graphics}
\usepackage[font=md,captionskip=8pt]{subfig}
\usepackage[usenames,dvipsnames]{color}
\usepackage[noadjust]{cite}
\usepackage{xcolor} 
\usepackage[utf8]{inputenc}
\usepackage{setspace}

\newcommand{\sg}{\sqrt{g}}    
\newcommand{\sqg}{\sqrt{g}}
\newcommand{\sqgh}{\sqrt{\hat g}}
\newcommand{\gh}{\hat g} 
\newcommand{\w}{w}

\newcommand{\q}{\alpha}
\newcommand{\h}{\varphi}

\newcommand{\tGamma}{\tilde\Gamma}
\newcommand{\xt}{(\vec x,t)}

\allowdisplaybreaks
\newcommand{\cL}{{\cal L}}

\newcommand{\cO}{{\cal O}}   

\newcommand{\cR}{{\tilde R}}

\newcommand{\cV}{{\mathcal V}}

\newcommand{\ra}{\rightarrow}

\newcommand{\be}{\begin{equation}}
\newcommand{\ee}{\end{equation}}
\newcommand{\bea}{\begin{eqnarray}}
\newcommand{\eea}{\end{eqnarray}}

\newcommand{\baa}{\begin{array}}
\newcommand{\eaa}{\end{array}}

\long\def\symbolfootnote[#1]#2{\begingroup
\def\thefootnote{\fnsymbol{footnote}}\footnote[#1]{#2}\endgroup}

\begin{document} 
\begin{flushright}
%{ \today}
\end{flushright}
\bigskip\medskip
\thispagestyle{empty}
\vspace{2cm}

\begin{center}
\vspace{0.5cm}

 {\Large \bf
 Gauging scale symmetry  and inflation:

\bigskip
 Weyl versus Palatini gravity}

\vspace{1.cm}

 {\bf D. M. Ghilencea}
\symbolfootnote[1]{E-mail: dumitru.ghilencea@cern.ch}

\bigskip
{\small Department of Theoretical Physics, National Institute of Physics
 
and Nuclear Engineering, Bucharest\, 077125, Romania}
\end{center}

\bigskip
\begin{abstract}
\begin{spacing}{1.07}\noindent
  We present a comparative study of inflation in 
  two theories of quadratic gravity   with  {\it gauged} scale symmetry:
1) the original Weyl  quadratic gravity and 2) the theory defined by a similar 
action but  in the  Palatini approach obtained by replacing  the Weyl
connection by its Palatini counterpart. These theories have  different vectorial
non-metricity induced by   the  gauge field ($\w_\mu$) of this symmetry.
Both theories have a novel  spontaneous breaking of
gauged scale symmetry, in the absence of matter,
where the necessary  scalar field is not added ad-hoc to
this purpose but is of geometric origin and part of the quadratic  action.
The  Einstein-Proca action (of $\w_\mu$), Planck scale  and metricity
emerge in the broken phase after $\w_\mu$ acquires  mass (Stueckelberg mechanism), then
decouples. In the presence of  matter ($\phi_1$), non-minimally coupled,  the scalar potential
is similar in both theories up to couplings and field rescaling. For small field
values the  potential is Higgs-like while for large fields  inflation is possible.
Due to their  $R^2$ term,  both theories have  a small tensor-to-scalar ratio
($r\sim 10^{-3}$), larger in Palatini case. For a fixed spectral index
$n_s$,  reducing  the  non-minimal coupling ($\xi_1$) increases $r$ 
which in Weyl theory is bounded from above by that of  Starobinsky inflation.
For a small enough $\xi_1\leq 10^{-3}$,  unlike the Palatini version,  Weyl theory
gives a dependence  $r(n_s)$   similar to that in Starobinsky inflation,
while also  protecting  $r$ against    higher dimensional operators corrections. 
\end{spacing}
\end{abstract}
\newpage

\section{Motivation}

In this work we present a comparative study of inflation in two
theories of  quadratic gravity  that have a {\it gauged}  scale symmetry
also known as  Weyl gauge symmetry.
This symmetry was first present  in the original Weyl quadratic gravity 
\cite{Weyl1,Weyl2,Weyl3}  (for a review \cite{Scholz})
that follows from  an  underlying
Weyl conformal geometry. This is relevant in early cosmology
when effective theories at short distances may become conformal.
Due to their symmetry, these theories  have no mass scales or dimensionful couplings - these
must be generated  by the vacuum expectations values (vev) of the fields
and this is the view we adopt here.

The first theory  is the original Weyl quadratic gravity 
revisited recently in \cite{Ghilen1,Winflation} with new results. This was in fact
the first gauge theory (of scale invariance)\footnote{
The literature  sometimes calls  Weyl gravity  the action of a 
Weyl-tensor-squared term (in Riemannian geometry). The Weyl action we discuss
is  the original one defined by Weyl geometry 
 \cite{Weyl1,Weyl2,Weyl3,Scholz} but without Weyl's unfortunate 
interpretation of its gauge boson as the real photon, and it  includes the aforementioned term.}.
The second theory~\cite{gPalatini}  has a similar action
but in the Palatini formalism \cite{E1,E2,c7,So1},
which means  replacing the Weyl  connection   by the Palatini connection.
In the  absence of matter the Lagrangian has the form
\bea
L_0=\sqrt{g} \Big\{ \,\frac{\xi_0}{4!} \, R(\tGamma,g)^2-\frac{1}{4\q^2}\,R_{[\mu\nu]}(\tGamma)^2\Big\}
\eea
where $\tGamma$ is the Weyl or Palatini connection,
respectively and $\xi_0$ and $\q$ are constants.
These  terms involve the (scalar and tensor) 
curvatures $R$ and $R_{\mu\nu}$  which are functions of  $\tGamma$; note
that $\tGamma$ is not determined by the metric $g_{\mu\nu}$.
 This is the  minimal action with such gauge symmetry.
More quadratic terms may be present  in both cases, see later.

  In both  theories the connection ($\tilde\Gamma$) 
  is   Weyl gauge invariant.
  Hence this  is not only a symmetry of the action,
   but also of the underlying geometry.
Both theories have  vectorial non-metricity  which is due to 
the dynamics of the gauge field $\w_\mu$ of  scale symmetry\footnote{If matter fields 
  are  present, they can also induce non-metricity.};  $\w_\mu$ is
dynamical since  for $\tGamma$ symmetric (which we assume to be the case)
the term  $R_{[\mu\nu]}^2\!\sim\! F_{\mu\nu}^2$
is just a  gauge kinetic term of $\w_\mu$.   And if  $\w_\mu$ is not 
dynamical it can easily be integrated out and both theories are Weyl integrable and metric
($\tilde\nabla_\mu g_{\mu\nu}=0$), see  e.g. \cite{Ghilen1,gPalatini}.
In both theories   % the {\it trace} of non-metricity   defines, as usual,
the Weyl gauge field is related to the {\it trace} of non-metricity:
$\w_\mu\propto  g^{\alpha\beta}\tilde\nabla_\mu g_{\alpha\beta}$
where $\tilde\nabla$ is computed with the Weyl or  Palatini connection.
The two theories have however a different  non-metricity {\it tensor}; % of these theories
this leads to different inflation predictions
that we discuss. We thus have a link between non-metricity and 
 inflation predictions.

Our  study of  these two theories  with  {\it gauged} scale symmetry is motivated\,by:

{\bf a)\,} In the {\it  absence of matter} both theories of quadratic gravity
have spontaneous breaking of this symmetry as it was shown for the first time 
in \cite{Ghilen1} for   Weyl quadratic theory
 and in  \cite{gPalatini} for the  Palatini theories.
In both cases the Einstein-Proca action of   $\w_\mu$ and the Planck scale 
 emerge in the broken phase, after $\w_\mu$ becomes {\it massive} by
``eating'' the Stueckelberg field (would-be Goldstone/dilaton); this is the field that ``linearises'' 
$R(\tGamma,g)^2$ in the action, as we shall detail.  After  $\w_\mu$ decouples
near the Planck scale $M\!\sim\!\langle{\rm dilaton}\rangle$, the  Einstein action 
is naturally obtained (together with metricity, see below)\footnote{This mechanism may be  more
 general and could apply to metric affine theories \cite{P1,P2,P3} (see also \cite{Del}).}.
  Thus, these theories provide a natural mass generation (Planck and $\w_\mu$ masses)
via a symmetry breaking mechanism.

The above result is important since it shows a new 
mechanism of spontaneous breaking of scale symmetry (in the absence of matter)
in which the necessary scalar field is
not added  ad-hoc 
to this purpose (as usually done);
instead, the Stueckelberg field is here  of {\it geometric} origin,
being ``extracted'' from the  $R(\tGamma,g)^2$  term. 
This situation is very different from  previous studies that
used instead e.g.  modified versions of Weyl action that were linear-only
in $R$  and/or  used additional matter field(s)  to
generate the Planck scale \cite{Dirac,Smolin,Cheng,Fulton,JW,Moffat1,Nishino,Oh,ghilen,Moffat2,Tann}.

{{\bf b)\,}
The breaking of Weyl gauge symmetry mentioned at a) is accompanied by
a change of the underlying geometry (connection). For example  in the Weyl theory
after $\w_\mu$ becomes massive it decouples,  the Weyl connection becomes
Levi-Civita,  thus the underlying Weyl geometry becomes Riemannian and  the theory 
becomes metric. A similar change of the underlying geometry happens in
the Palatini case. Hence,  the breaking of the Weyl gauge symmetry shown in 
\cite{Ghilen1,gPalatini} is not the result of a
mere choice of a gauge (as it happens in Weyl or conformal theories with no Weyl gauge field), 
but is more profound: it is accompanied by both  a Stueckelberg mechanism (as mentioned)
and by transformations at a  geometric level.}

{\bf c)\,} In both Weyl and Palatini theories  $\w_\mu$
has a large mass ($\sim M$) \cite{Ghilen1,gPalatini} so the associated non-metricity scale
is very high;  hence,
non-metricity effects are suppressed by $M$. One thus avoids  long-held criticisms 
\cite{Weyl1} that had assumed a {\it massless} $w_\mu$ (implying metricity violation
at low scales or path dependence of clock's rates/rod's length,
in contrast to experience \cite{Latorre}).

{\bf d)\,} If  matter is present e.g. a Higgs-like scalar  is non-minimally coupled
 to $R(\tGamma,g)$, Weyl and Palatini theories have successful inflation, in addition to
 mass  generation.
The main  goal of this work is to investigate comparatively
 their inflation predictions.  We give new results in Section~\ref{sec3},
 such as  the dependence $r(n_s)$
of the tensor-to-scalar ratio $r$ on the spectral index $n_s$ in
Weyl and Palatini cases and their relation to Starobinsky inflation~\cite{Sta}.

{\bf e)\,}  The Standard Model (SM) with a vanishing Higgs mass has a  Weyl gauge symmetry. 
It is well-known that the fermions and gauge bosons do not couple 
to the gauge field $\w_\mu$ \cite{Kugo} but scalars (Higgs) have couplings to $\w_\mu$.
Having seen that  $\w_\mu$ is  massive  \cite{Ghilen1,gPalatini}
it is  worth studying the  SM in Weyl  quadratic  gravity or its Palatini version\footnote{
For the SM Lagrangian in Weyl quadratic gravity 
see  \cite{Ghilen1} (second reference, Section 1.7) and \cite{Kugo,Moffat1,ghilen}.}. 
If the gauged scale symmetry is relevant for the mass hierarchy problem, it is intriguing 
that only the Higgs field 
couples directly to the gauge boson $\w_\mu$ of scale symmetry.

{\bf f)\,} $\w_\mu$ is a dark matter candidate \cite{Tang} and,  being part of $\tGamma$, 
it could give  a {\it geometric} solution to the dark matter problem.
 This  brings together physics beyond SM and gravity.

{\bf g)\,} The models with {\it gauged} scale symmetry 
do not have the unitarity issue (negative kinetic term)
present in {\it local} scale invariant Lagrangians (without $\w_\mu$),
when generating the  Einstein action from such Lagrangians: 
$L=(-1/12)\sqg\, [\phi^2 R+ 6 (\partial_\mu\phi)^2]$.
See \cite{Oh} for a  discussion on  this issue in
local scale invariant models\footnote{
Avoiding unitarity violation in local scale invariant cases
may require $\phi$ have an 
imaginary vev \cite{tH0,tH3} but then the associated  conformal transformation
involving $\Omega^2\!\propto\!\phi^2$ seems to change the overall metric signature.}
 \cite{tH0,tH3,Turok,Bars1,Bars2,Kallosh}.
In a {\it gauged} scale invariant model
this negative kinetic term is cancelled  and  $\phi$ is ``eaten'' 
by $\w_\mu$ which acquires mass \cite{Ghilen1,gPalatini} 
\`a la Stueckelberg \cite{ST,P2} and  decouples,
to recover the  Einstein action and gauge.

{\bf h)\,} In the gauged \cite{Ghilen1,gPalatini} and global \cite{FRH1,FRH2,higgsdilaton} cases
 there is an associated non-zero conserved current, unlike in
 some  local scale invariant models where this current is trivial \cite{J1,J2}.

{\bf i)} A {\it gauged} scale symmetry  seems stable under
 black-hole physics unlike a global one\cite{Kalloshp}, so it
 is preferable when building models that
include gravity. Global  models are easily made
 gauged  scale invariant  by replacing 
 their Levi-Civita connection by e.g.\,Weyl connection.
 The theories discussed can give a gauged scale invariant version of
 Agravity global model\cite{agravity1,agravity2}.

{ {\bf j)} Another motivation to study  theories  with Weyl gauge symmetry
  is their geodesic completeness, as  emphasized in \cite{Oh} and summarised here.
  In conformal invariant theories geodesic completeness can be achieved
  without the Weyl vector presence,  in  the (metric)  Riemannian universe;
  there,  geodesic completeness or incompleteness
  is related to a specific gauge choice
  (with singularities due to an unphysical conformal frame) \cite{narlikar,modestoJCAP,modesto1605,Rachwal}.
 But Weyl gauge symmetry is more profound and complete: it is more than a symmetry of the action
 since, (unlike in  conformal/Weyl invariant theory  with no $\w_\mu$),
 it is  also a  symmetry of the underlying geometry (of connection $\tilde\Gamma$).
 The geodesics  are then determined by the affine structure and
 differential geometry {\it demands}  the existence of the Weyl gauge field \cite{Ehlers}
  for the construction of the affine connection, % of Weyl geometry
  because this ensures that geodesics are  invariant (as required on  physical arguments).
  After the breaking, $w_\mu$  decouples,
  see b) above,  and we return to
  Riemannian geometry with geodesics  given by extremal proper time condition\footnote{
    Since the  Weyl gauge field brings in non-metricity, geodesic completeness
    is related to non-metricity.}.
} 

The above  arguments, a) to j), 
motivated our interest in  theories beyond Standard Model (SM)  with  Weyl gauge symmetry.
Section~\ref{sec2}  reviews the two theories, showing their similarities
and differences,  see \cite{Ghilen1,gPalatini} for technical details. 
Section~\ref{sec3} studies comparatively their inflation predictions.
The Appendix has technical details and an application to inflation.

\medskip
\section{Weyl versus Palatini quadratic gravity}\label{sec2}

\subsection{The symmetry}\label{2.1}

Consider a Weyl local scale transformation $\Omega(x)$
 of the metric $g_{\mu\nu}$ and of a scalar field $\phi_1$ \footnote{
Our conventions are those in the Appendix of \cite{book} with  metric  (+,-,-,-), \,\,
 and $g\equiv \vert \det g_{\mu\nu}\vert$.} 
%\medskip
\bea
\label{cr}
\hat g_{\mu\nu}=\Omega^2 g_{\mu\nu},\quad
\sqgh=\Omega^4\sqg,\quad
\quad \hat\phi_1=\frac{1}{\Omega}\,\phi_1.
\eea
%\medskip\noindent
To this geometric transformation one associates a Weyl gauge field $\w_\mu$
that transforms as 
%\medskip
\bea\label{s2}
\hat \w_\mu=\w_\mu-\partial_\mu \ln \Omega^2.
\eea
%\medskip\noindent
Eqs.(\ref{cr}), (\ref{s2}) define a gauged scale transformation.
The symmetry is  a gauged dilation group  isomorphic to $R^+$ (non-compact). It differs
from internal gauge symmetries, since $\Omega$ is real. 

What is the  relation of the Weyl field to the underlying geometry which is defined by
 $g_{\mu\nu}$ and $\tGamma$? One can define $\w_\mu$ via the 
non-metricity, but it is more intuitive to define 
 $\w_\mu$  as a measure of the deviation of  (the trace of)  $\tGamma$ 
 from the Levi-Civita connection:
\medskip
\bea\label{trace}
\w_\mu=(1/2)\,(\tGamma_\mu-\Gamma_\mu(g)),
\eea

\medskip\noindent
with a notation $\tGamma_\mu=\tGamma_{\mu\nu}^\nu$ and $\Gamma_\mu=\Gamma_{\mu\nu}^\nu(g)$.
 $\Gamma_{\mu\nu}^\alpha(g)$ is the Levi-Civita connection for  $g_{\mu\nu}$  while 
$\tGamma_{\mu\nu}^\alpha$ is the connection in either Weyl or Palatini gravity.
We assume a symmetric connection $\tGamma_{\mu\nu}^\alpha=\tGamma_{\nu\mu}^\alpha$ 
(no torsion). Note  that $\w_\mu$ is  a vector under coordinate 
transformation ($\tGamma_\mu$ and $\Gamma_\mu$ are not).
 Finally, $\tGamma_{\mu\nu}^\alpha$ and in particular 
$\tGamma_\mu$  is invariant under (\ref{cr}), (\ref{s2}),
 in both Weyl and Palatini gravity (see also  the Appendix).  
To check this invariance use  (\ref{s2}) in (\ref{trace}) and that 
 $\Gamma_\mu(g)=\partial_\mu\ln\sqg$;  then $\Gamma_\mu(\hat g)=\partial_\mu\ln (\sqg \,\Omega^4)$.
The change of the metric is compensated by that of $\w_\mu$, leaving
$\tGamma_\mu$ invariant.

\subsection{The Lagrangian:  Weyl versus Palatini}

Consider next a Lagrangian with gauged scale invariance 
for a  scalar field with non-minimal coupling,
 in Weyl  and Palatini  quadratic  gravity.
The analysis being  similar, we present simultaneously both Weyl 
and Palatini theories. The main difference between them  is in the
coefficients $\tilde\Gamma_{\mu\nu}^\alpha$ which we do not need to specify right now. 
Consider then a (Higgs-like) scalar $\phi_1$ with non-minimal coupling  $\xi_1\!>\! 0$:
\medskip
\bea\label{ll1}
L=\sg\,\Big[\,
\frac{\xi_0}{4!}\,R(\tGamma,g)^2 
 - \frac{1}{4 \q^2}\, F_{\mu\nu}(\tGamma)^2
- \frac{1}{12} \,\xi_1\phi_1^2\,R(\tGamma,g) 
+\frac12 \,(\tilde D_\mu\phi_1)^2 - \frac{\lambda_1}{4!}\phi_1^4\Big],
\eea

\medskip\noindent
with a scalar curvature $R(\tilde\Gamma,g)$ 
which depends on the Weyl or Palatini connection $\tilde\Gamma$:
\medskip
\bea\label{e02}
R(\tGamma,g)=g^{\mu\nu}\,R_{\mu\nu}(\tGamma),\qquad
R_{\mu\nu}(\tGamma)=\partial_\lambda\tilde\Gamma^\lambda_{\mu\nu}-
\partial_\mu \tGamma^\lambda_{\lambda\nu}
+\tGamma_{\rho\lambda}^\lambda \tGamma^\rho_{\mu\nu}-\tGamma^\lambda_{\rho\mu}
\tGamma^\rho_{\nu\lambda}.
\eea

\medskip\noindent
 $\tGamma$ is invariant under (\ref{cr}), (\ref{s2})
so $R_{\mu\nu}(\tGamma)$ is invariant; $R(\tGamma,g)$
 transforms covariantly, (\ref{RR})
\medskip
\bea\label{r}
\hat R(\tGamma,\hat g)=(1/\Omega^2)\, R(\tGamma,g).
\eea

\medskip\noindent
With  (\ref{r}), the first and third  term in $L$ are
 invariant under (\ref{cr}), (\ref{s2}). 

Further, the second term in $L$ is a  gauge kinetic term of $\w_\mu$ and involves 
\medskip
\bea\label{fmunu}
F_{\mu\nu}(\tilde\Gamma)=
 \tilde\nabla_\mu \w_\nu -\tilde\nabla_\nu \w_\mu
=\!\partial_\mu \w_\nu-\partial_\nu \w_\mu
=(\partial_\mu\tGamma_\nu-\partial_\nu\tGamma_\mu)/2.
\eea

\medskip\noindent
 $\tilde\nabla$ is defined by  $\tGamma$
and in the second step  we used that $\tGamma$ is symmetric.
From  (\ref{fmunu})  $F_{\mu\nu}$ is invariant under (\ref{cr}), (\ref{s2}),
and one verifies that the second term in $L$ is also invariant under these transformations.
Since  $F_{\mu\nu}(\tGamma)^2=\,R_{[\mu\nu]}(\tGamma)^2$ 
where  $R_{[\mu\nu]}\equiv (R_{\mu\nu}-R_{\nu\mu})/2$,
 a gauged scale symmetry is naturally present
in the Palatini version of $R^2+R_{[\mu\nu]}^2$  gravity.

The Weyl-covariant derivative of $\phi_1$ in $L$ and its 
transformation under (\ref{cr}), (\ref{s2}) are
\bea\label{ttt}
\tilde D_\mu\phi_1=(\partial_\mu-1/2\,\w_\mu)\,\phi_1,\qquad
\hat{\tilde D}_\mu\hat \phi_1=(1/\Omega)\, \tilde D_\mu\phi_1.
\eea

\medskip\noindent 
Therefore $\phi_1$ is charged under the Weyl gauge symmetry.
With (\ref{ttt}) one checks that the kinetic term of $\phi_1$ is invariant under
 (\ref{cr}), (\ref{s2}). Finally, $\lambda_1\phi_1^4$ is the 
only potential term allowed by symmetry, so  the entire $L$ is invariant.

In the {\it absence of matter}\, ($\phi_1$), $L$ contains the first two terms only,
giving the minimal action of the original Weyl quadratic gravity or its Palatini version; 
both actions have gauged scale symmetry and, after spontaneous breaking of this symmetry, 
one  obtains the Einstein-Proca action for $\w_\mu$, see \cite{Ghilen1,gPalatini}.
If only the first term is present in $L$, both theories 
are Weyl integrable (metric) and  Einstein action is obtained with
 a positive cosmological constant.

Returning to $L$, we replace the first term  in (\ref{ll1})  by
 $\xi_0 R(\tGamma,g)^2\ra -\xi_0\,(2\, \phi_0^2\, R(\tilde\Gamma,g)+\phi_0^4)$
 where $\phi_0$ is an auxiliary scalar;  using the equation of motion of $\phi_0$ (of solution 
$\phi_0^2=-R$) recovers onshell the term $\xi_0\,R^2$  in (\ref{ll1}).
 This gives a classically equivalent  $L$, {\it linear} in $R$
\medskip
\be\label{ll1prime}
L=\sg\,\Big[\,
- \frac{1}{2} \,\rho^2\,R(\tGamma,g)
- \frac{1}{4 \q^2}\,F_{\mu\nu}^2 
+
\frac12 \,(\tilde D_\mu\phi_1)^2
-\cV(\phi_1,\rho)
\Big],
\ee
where
\be\label{toE}
\cV(\phi_1,\rho)=\frac{1}{4!}\,\Big[\,
\frac{1}{\xi_0} \big(6\rho^2-\xi_1\phi_1^2\big)^2
+
\lambda_1\phi_1^4\,\Big],\qquad\text{and}\qquad
\rho^2=\frac16\,\big(\xi_1\phi_1^2+\xi_0\phi^2_0).
\ee

\medskip\noindent
We further replaced $\phi_0$ by  radial direction $\rho$ in field space, so
our new fields are now $\{\rho, \phi_1\}$.

$L$ has similarities to a global
scale invariant Higgs-dilaton model, eqs. (2.9), (2.10) of \cite{higgsdilaton} 
also \cite{CW1,CW2};   $\phi_0$ has a large coupling 
 ($\xi_0\!>\!1$) to $R$ since the $R^2$  term has a perturbative coupling $1/\sqrt{\xi_0}\!<\!1$
and this corresponds to a  Higgs  of non-minimal  coupling  $\xi_h\!>\!1$ in \cite{higgsdilaton}.

The action in  (\ref{ll1prime}) depends on $\tGamma$ through its first three terms.
We have  two cases:

{\bf a).\,\,\,}  In Weyl quadratic gravity, $\tGamma$ is determined by
$g_{\mu\nu}$ and the gauge field  $\w_\mu$, see its expression in  eq.(\ref{o}) in the Appendix.
Using this one replaces  the scalar curvature in (\ref{ll1prime}) in terms of the Ricci scalar
of Riemannian geometry, eq.(\ref{Wcurvature}).
 The result is eq.(\ref{ll2}) below.

{\bf b).\,\,\,} In  Palatini  gravity,  $\tilde\Gamma$
is simply determined by its equation of motion from the action in (\ref{ll1prime}). 
After solving   this  equation \cite{gPalatini}, we obtain the connection
shown in
eq.(\ref{Vm1}) in the Appendix; $\tGamma$ differs from that in Weyl case, due to 
different non-metricity (accounted for by $\gamma$ in eq.(\ref{ll2})).
With this  $\tGamma$, one computes the scalar curvature,as usually  done 
eq.(\ref{Pcurvature}). Replacing this curvature back  in action 
(\ref{ll1prime}) one finds again $L$ below (for $\tGamma$ onshell):
\medskip
\be\label{ll2}
L=\sqg \Big\{
\frac{-1}{2} \Big[\rho^2 R(g) + 6 (\partial_\mu\rho)^2\Big] + 3 \gamma
\,\rho^2 (\w_\mu-\partial_\mu\ln\rho^2)^2
-\frac{1}{4 \q^2} \,F_{\mu\nu}^2 +\frac12 (\tilde D_\mu\phi_1)^2 - \cV(\phi_1,\rho)\Big\}
\ee
\bea%\label{eqf}
{\rm where}\qquad \gamma=1/4\,\,\, \rm{for\, Weyl\,\, case};
\quad\qquad
 \gamma=1\,\,  \rm{for\, Palatini\,\, case.}
\eea
$R(g)$ is the Ricci scalar for the metric $g_{\mu\nu}$. 
This is a metric 
formulation equivalent to the initial Lagrangian eq.(\ref{ll1}),
 invariant under transformations (\ref{cr}), (\ref{s2}); under these
$\ln\rho$ transforms with a shift,
 $\ln\rho\!\ra\!\ln\rho\!-\!\ln \Omega$, so $\ln\rho$  acts like  a
would-be Goldstone (``dilaton''), see later.

\subsection{Einstein-Proca action as a broken phase of Weyl or Palatini  gravity }

Since $L$ has a gauged scale symmetry, we should ``fix the gauge''. 
We choose  the Einstein gauge corresponding to constant  $\rho$; this is obtained 
by using a  transformation (\ref{cr}),(\ref{s2}) of a particular
$\Omega\!=\!\rho/\langle\rho\rangle$ which is $\rho-$dependent and 
sets  $\hat\rho$ to a {\it constant} $\hat\rho=\langle\rho\rangle$, and so
introduces a mass scale. In terms of  new variables (with a hat) eq.(\ref{ll2}) becomes 
\medskip
\be\label{W3}
L=
\sqrt{\hat g}\, \Big\{-\frac{1}{2} M^2\,R(\gh) +3 \gamma\, M^2 \hat \w_\mu\hat \w^\mu
-\frac{1}{4 \q^2}  \hat F_{\mu\nu}^2 +\frac{1}{2}(\hat{\tilde D}_\mu\hat\phi_1)^2
-\cV(\hat\phi_1,M) \Big]\Big\},
\ee

\medskip\noindent
 with $R(\hat g)$ the Ricci scalar for metric $\hat g_{\mu\nu}$, 
 ${\hat{\tilde D}}_\mu\hat\phi_1=(\partial_\mu-1/2\,\,\hat \w_\mu )\hat\phi_1$
 and with $\nabla_\mu \hat\w^\mu=0$;  we denoted  $M=\langle\rho\rangle$ 
 which we identify with the Planck scale. The potential  now depends
on $\hat\phi_1$ only, see (\ref{toE}).
This is the Einstein-Proca action for $\hat\w_\mu$: 
this field  has become massive of mass $m_\w^2=6 \alpha \,\gamma\, M^2$
 by absorbing the derivative of
the Stueckelberg (would-be ``dilaton'') field $\partial_\mu\ln\rho$; then the {\it radial} direction
 in field space ($\rho$) is not present anymore in the action. 
This is a spontaneous breaking of Weyl gauge symmetry; the number $n$ of  degrees 
of freedom other than the graviton ($n=3$) is conserved during this breaking:
the initial massless scalar $\rho$ and massless vector  $\w_\mu$ are replaced by 
a massive gauge field $\w_\mu$.

Note that in the absence of matter  ($\phi_1$), the Stueckelberg field needed for breaking
becomes  $\ln\rho\propto \ln\phi_0$ and has a pure geometric origin,
being simply  ``extracted'' from the quadratic curvature term $R^2(\tGamma,g)$ in
the initial,  symmetric  action.
Therefore, one does not need to add this scalar field ad-hoc as usually done
to this purpose,  and the breaking and mass generation ($m_w$, Planck scale)
takes place even in the absence of matter \cite{Ghilen1,gPalatini}.
Finally, unless one is tuning the coupling $\alpha$ to small values,
the   mass of  $\hat\w_\mu$ is near  the Planck scale\footnote{This 
is  preferable, since then one avoids metricity violation 
 below the Planck scale (due to a lighter $\w_\mu$). 
Current non-metricity lower bounds could be as low as TeV \cite{Latorre} but are 
 model dependent.}.

\subsection{Scalar potential}

To obtain a  standard  kinetic term for $\hat\phi_1$,
similar to the   ``unitarity  gauge'' in the electroweak case, we  remove the
coupling  $\hat \w^\mu \partial_\mu \hat\phi_1$ from the term  $(\hat{\tilde D}_\mu\hat\phi_1)^2$ in
(\ref{W3})
by a field redefinition 
\medskip
\bea\label{tt}
\hat\w_\mu \ra  \hat\w_\mu+
 % {\hat \w_\mu^\prime} =\hat \w_\mu - 
\partial_\mu \ln \cosh^2 \Big[\frac{\h}{2 M\sqrt{6\gamma}}\Big],\qquad
\hat \phi_1
\ra 2 M\sqrt{6\gamma}\,\sinh\Big[\frac{\h}{2 M\sqrt {6\gamma}}\Big]
\eea

\medskip\noindent
In terms of the new fields  eq.(\ref{W3}) becomes
\medskip
\be\label{twp}
L=\sqrt{\hat g}\, \Big\{
-\frac{1}{2} M^2 R (\hat g)
+ 3\gamma  M^2 \cosh^2 \Big[\frac{\h}{2 M\sqrt {6\gamma}}\Big] 
\,\hat \w_\mu\hat \w^{\mu}
-\frac{1}{4 \q^2} \hat F^{ 2}_{\mu\nu} % F^{\prime\, 2}_{\mu\nu}  
+\frac{\hat g^{\mu\nu}}{2}\partial_\mu\h \partial_\nu \h
- V(\varphi)
\Big\},
\ee

\medskip\noindent
which is ghost-free and
\medskip
\bea\label{scalars2}
V(\varphi)=
V_0\,
\Big\{
 \Big[
1-(4 \gamma)\,\xi_1\,\sinh^2 \frac{\h}{2 M\sqrt{6\gamma}}\Big]^2+
(4 \,\gamma)^2\,\lambda_1  \xi_0 \,\sinh^4\frac{\h}{2 M\sqrt {6\gamma}}
\,\Big\},\quad 
V_0\equiv\frac{3}{2} \frac{M^4}{\xi_0}.
\eea

\medskip\noindent
Lagrangian (\ref{twp}) describes Einstein gravity,
a scalar field $\varphi$ with canonical kinetic term
and potential (\ref{scalars2}) that is $\gamma$-dependent,
and a massive Proca field  ($\hat\w_\mu$)  that decouples near the Planck scale $M$.
To make obvious the mass term of $\w_\mu$ in (\ref{twp})
use that  $\cosh^2x=1+\sinh^2x$.
Eqs.(\ref{twp}), (\ref{scalars2})  can be extended to more scalar fields,
 see second reference in \cite{Ghilen1} (eq.24). % for the  Weyl case.

For {\it small field} values $\h\ll M$, the potential in (\ref{scalars2})  becomes 
(recall that $M=\langle\rho\rangle$):
\medskip
\bea\label{fghj}
V(\varphi)
=\frac{3 \langle\rho\rangle^4}{2\xi_0}
-\frac{
1}{2} \,\frac{\xi_1}{\xi_0}\,\langle\rho\rangle^2\,\h^2
+
\frac{ 
1}{4!}\Big[ \lambda_1 + \frac{\xi_1}{\xi_0}\Big(\xi_1-\frac{1}{6\gamma}\Big)
\Big]\,\h^4 +\cO(\h^6/\langle\rho\rangle^2).
\eea

\medskip
In this case the  potential is similar in  Weyl and Palatini cases,
up to a small $\gamma$-dependence of the quartic coupling,
negligible  for (ultra)weak couplings  $\xi_1 /\xi_0\ll 1$;
in this case also the quadratic coupling is suppressed
(recall  the perturbative couplings are $1/\sqrt\xi_0\!<\!1$ and $\xi_1\!<\!1$).

If we identify $\h$ with the Higgs field, we have
electroweak symmetry breaking, since $\xi_1>0$.
For a classical hierarchy $\xi_1/\xi_0\ll 1$ one may be able
to tune the mass of $\h$ near  the electroweak scale $m^2=(\xi_1/\xi_0) \langle\rho\rangle^2$.
Gravitational corrections to  $\lambda_1$ may be negative
but there is no instability: the exact form of $V(\varphi)$ is  positive,
even if the self-coupling $\lambda_1=0$!

 For {\it large} $\varphi$ the potential  is different  in Weyl and Palatini cases
due to a different $\gamma$. 
This potential changed from initial (\ref{ll1}) to (\ref{scalars2})
following two steps: the ``linearisation'' of the $R^2$ term by  $\phi_0$ 
that induced the $\phi_0^4$ term,  then transformation (\ref{tt}) which decoupled the 
(trace of)  the connection from $\partial_\mu\phi_1$ and brought 
the presence of $\gamma$ i.e. non-metricity dependence.

 \begin{figure}[t!]
 \begin{center}
 \includegraphics[height=0.39\textwidth]{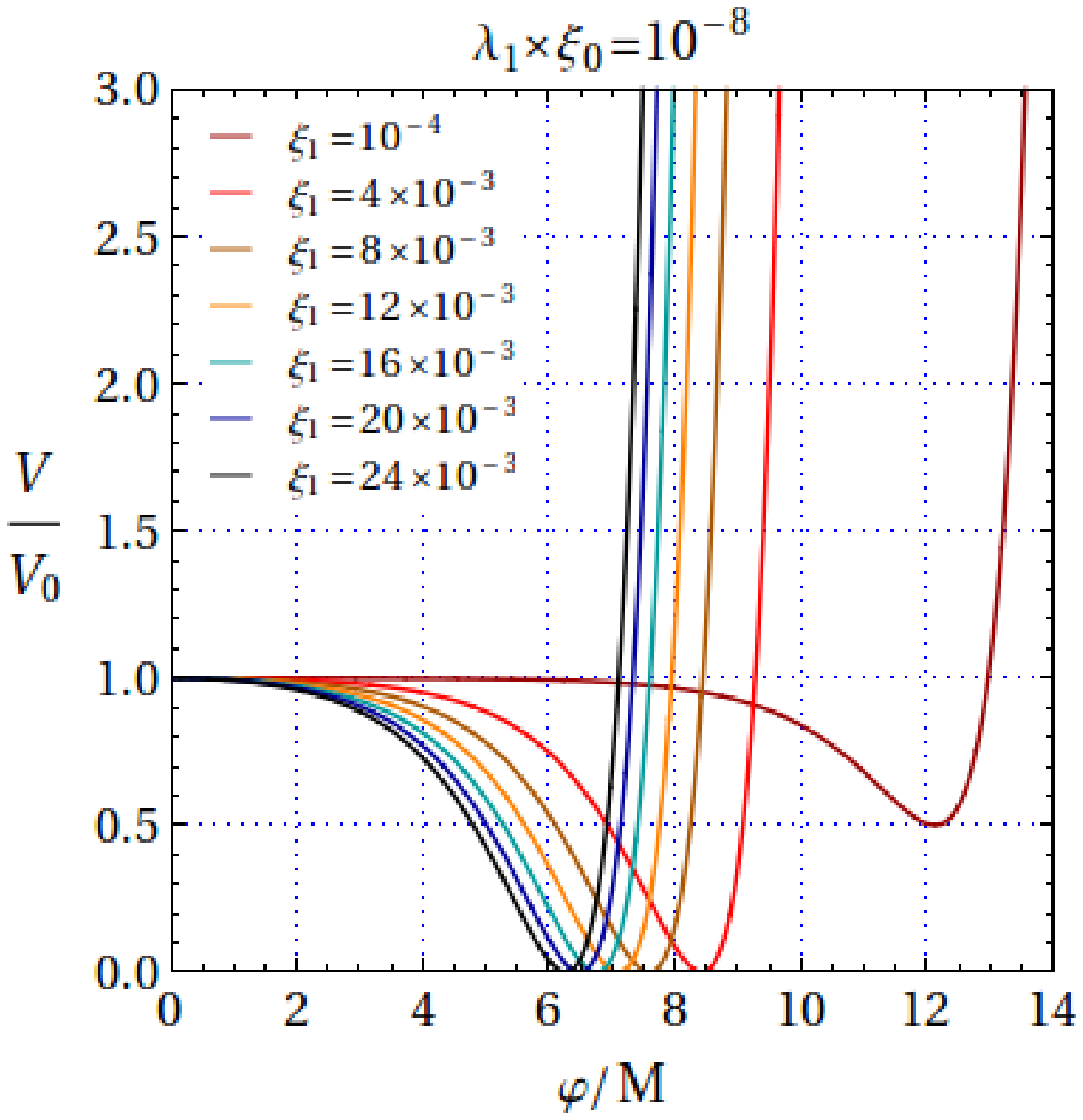}\qquad
 \includegraphics[height=0.39\textwidth]{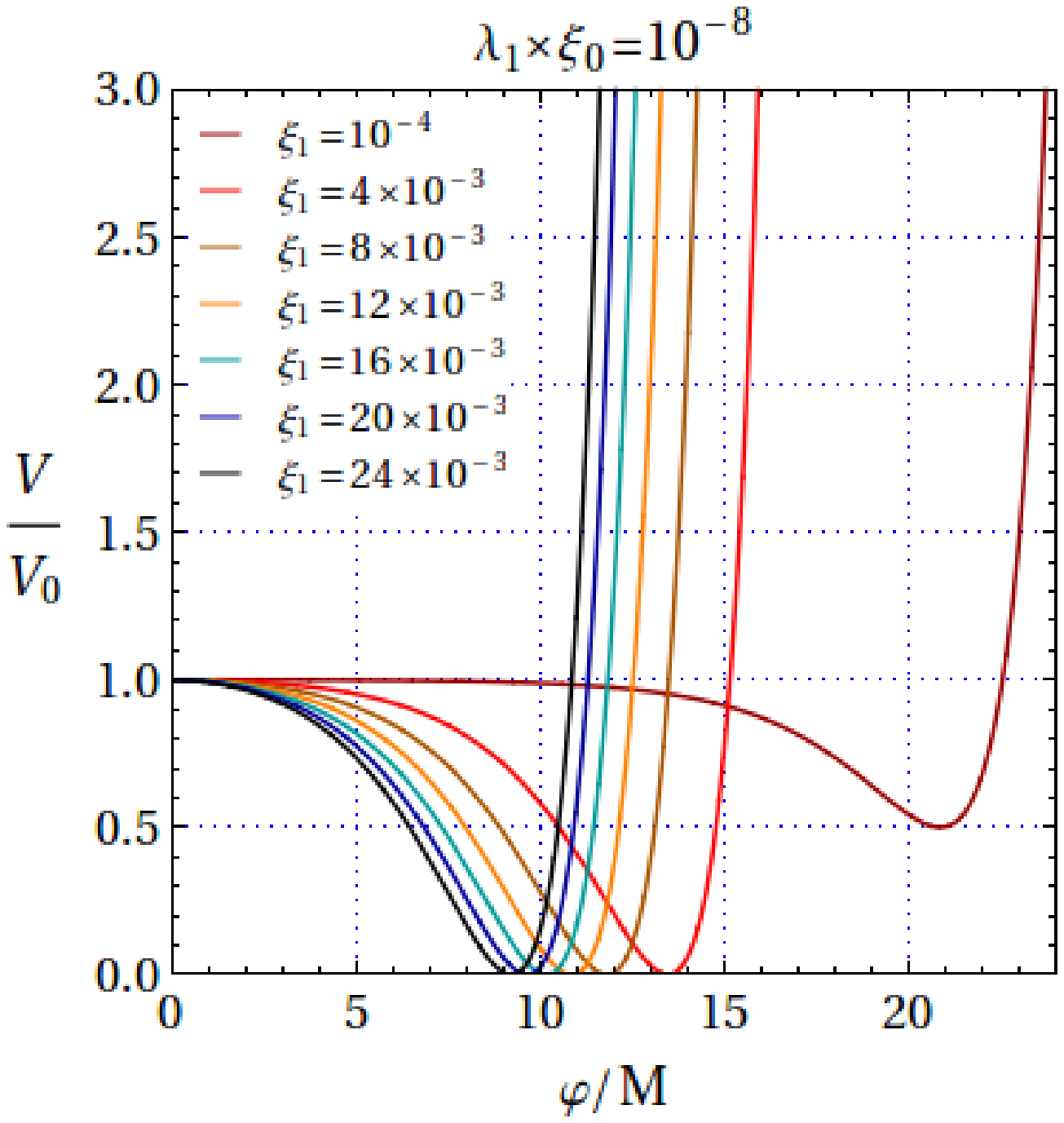}\qquad
 \end{center}
\vspace{-0.8cm}
\begin{center}
\quad
\includegraphics[height=0.39\textwidth]{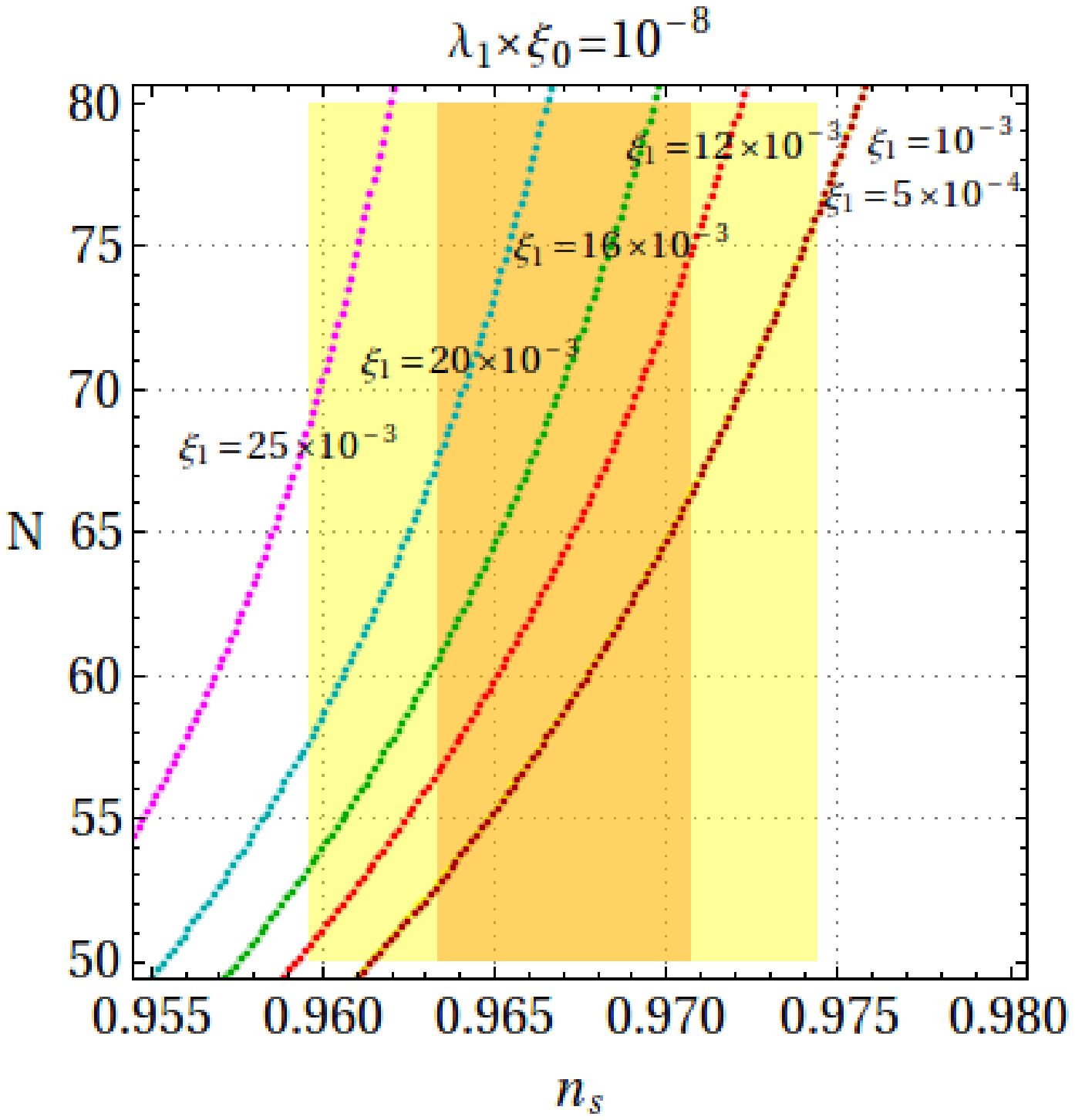}\quad\quad\,\,\,
\includegraphics[height=0.39\textwidth]{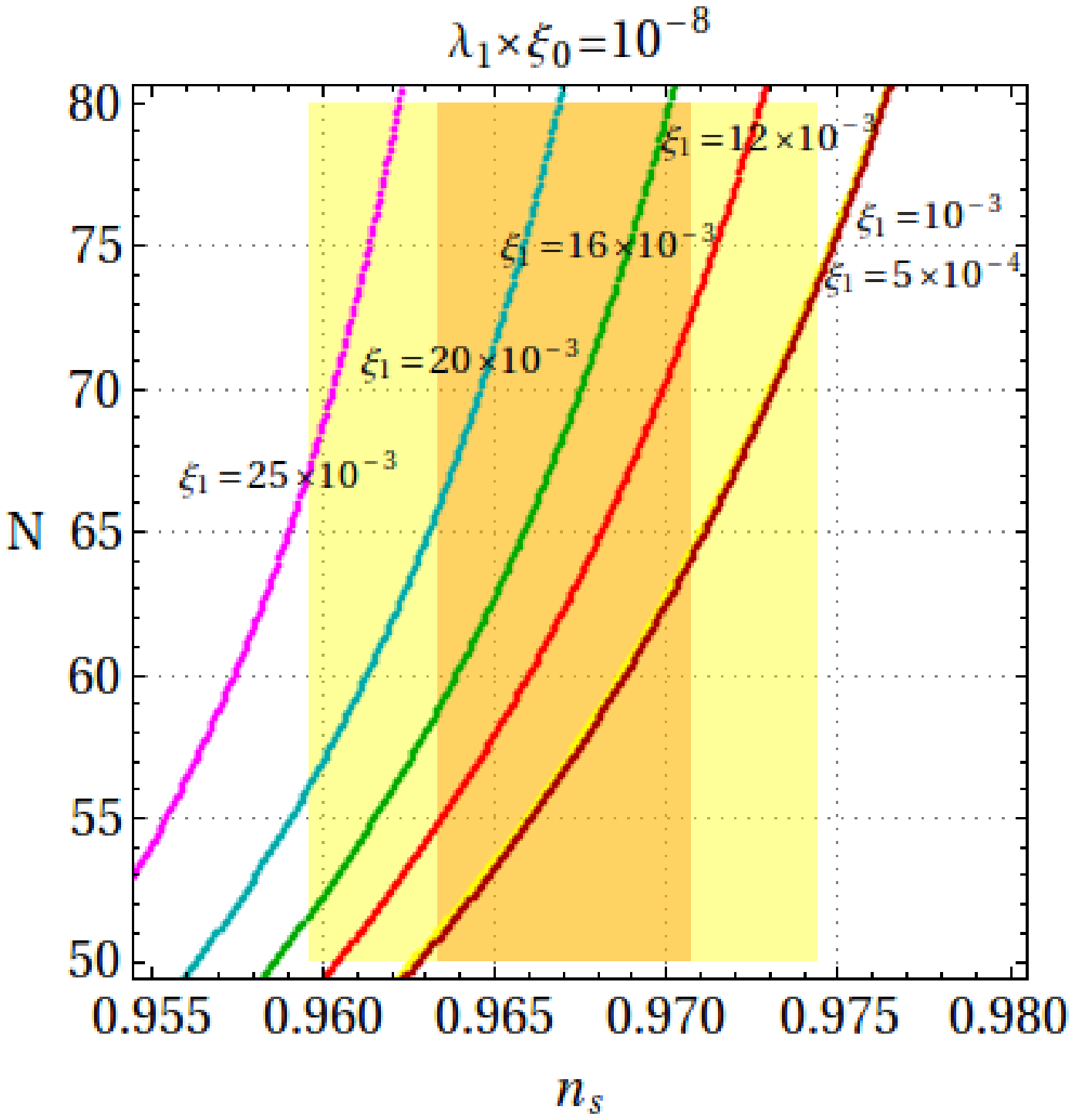}
\end{center}
\vspace{-0.75cm}
\begin{center}
\includegraphics[height=0.39\textwidth]{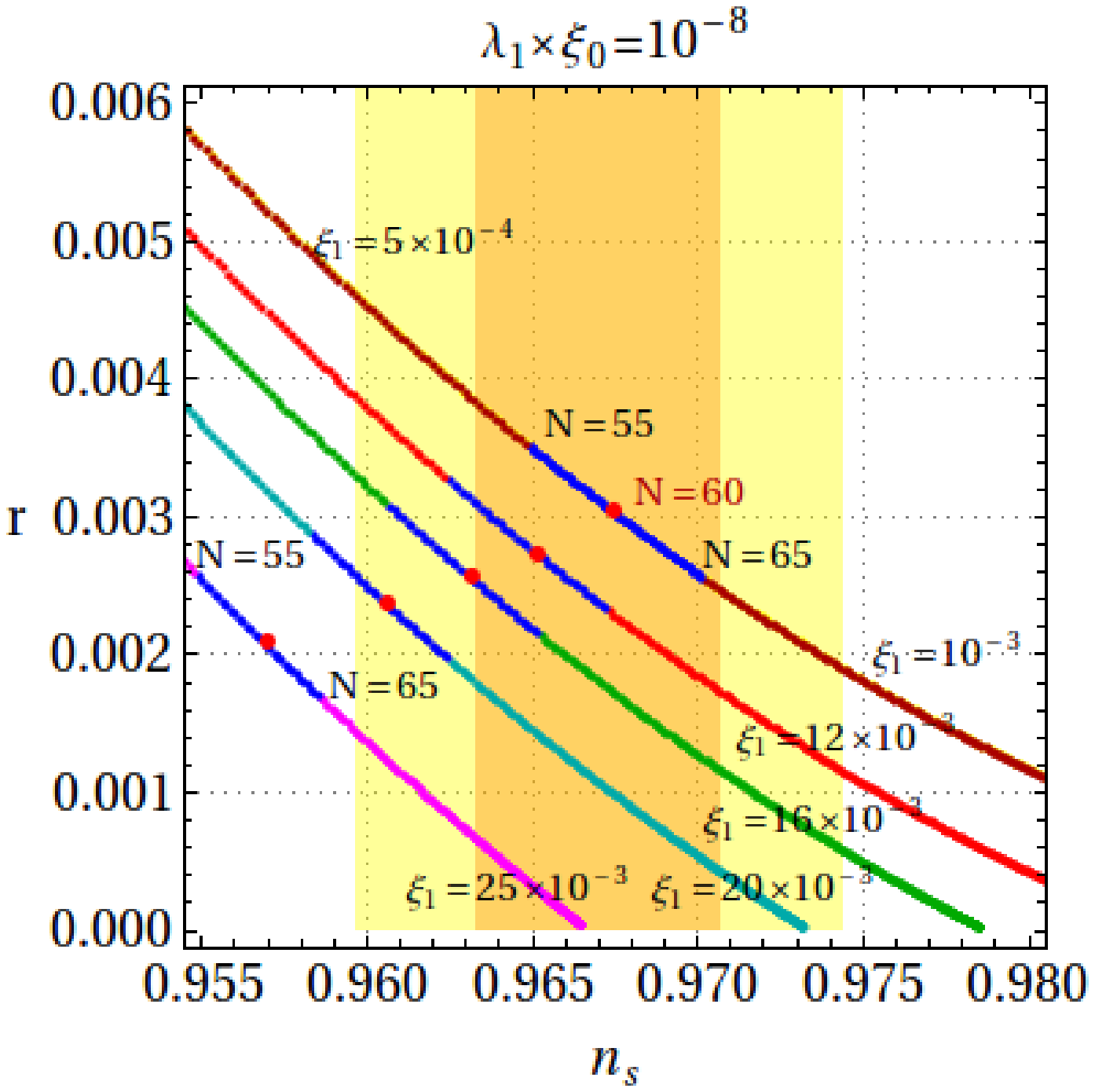}\quad\quad
\includegraphics[height=0.39\textwidth]{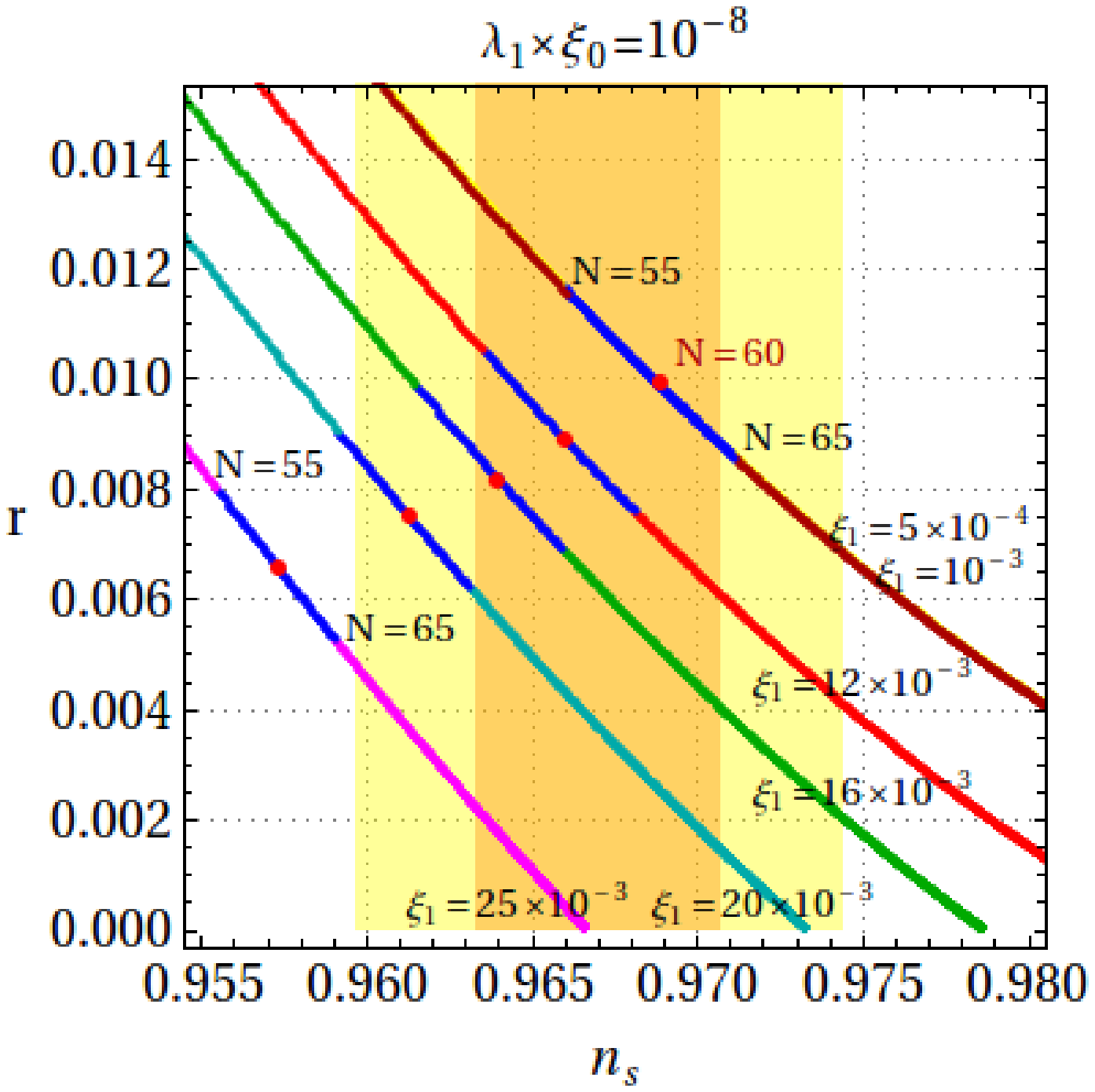}
\end{center}
\vspace{-0.25cm}
\caption{\small
{\bf Left column:}  Weyl inflation plots;  {\bf Right column:} Palatini inflation plots.
 All figures have $\lambda_1 \xi_0=10^{-8}\leq \xi_1^2$;
Top plots: the  potential $V/V_0$ in terms of $\varphi/M$ for different $\xi_1$;
larger $\xi_1$ moves the curves to the left; larger $\lambda_1\xi_0$ lifts the minimum
of the rightmost curves, see eq.(\ref{scalars2});
middle plots: the plots $(n_s, N)$ for different $\xi_1$;
bottom plots: the plots $(n_s,r)$ for different $\xi_1$; curves of $\xi_1=5\times 10^{-4}$ and $10^{-3}$
 are degenerate;
 along these curves the value of $N$ changes:
  the blue line segment has points of $55\leq N\leq 65$; red dots have
 $N=60$.  The yellow (orange) area
corresponds to the measured value of
$n_s$ at $95 \%$ CL $(68 \%)$, respectively.}
\label{fig01}
\end{figure}
%\medskip

\section{Inflation}\label{sec3}

\subsection{Weyl versus Palatini}\label{rs3}

We can now use Lagrangian (\ref{twp}) and potential $V(\varphi)$ of (\ref{scalars2})
to study inflation with $\varphi$ as the inflaton
and compare its  predictions for the  Weyl ($\gamma=1/4$)  and Palatini ($\gamma=1$) cases.
For a previous study of inflation in the  Weyl case, see\footnote{
For inflation in related Palatini models but without Weyl gauge symmetry, see
 \cite{II1,II2,Sh11,II3,II4,II5,II6,II7,Palatini1,Palatini2,Antoniadis,Antoniadis2,Gialamas,Gialamas2,Das}.}
\cite{Ross,Winflation}.
Lagrangian (\ref{twp}) describes  a single scalar field in Einstein gravity and the
usual formalism for a single-field inflation can  be used. However, 
notice there exists a coupling of $\varphi$ to the Weyl  field $\hat\w_\mu$, 
the second term in  (\ref{twp}). Hence, we must first show that this coupling and
$\hat\w_\mu$  do not affect inflation by~$\varphi$.

Firstly, we do not consider here the possibility of
the Weyl vector field itself as the inflaton\footnote{
  Inflation by vector fields was suggested in \cite{Ford,Lidsey}.}
since it could induce a  substantial large-scale
anisotropy \cite{Mukh} which would be in conflict with CMB isotropy.
The anisotropy is obvious in  the
stress-energy tensor contribution of $\hat\w_\mu$ which is not diagonal.
This issue  can be avoided if one considers  a large number of randomly oriented vector fields
or a triplet of mutually orthogonal vector fields
\cite{Mukh},  however this is not possible in the current fixed setup.

Secondly, one may ask if the Weyl field  could play the role of a curvaton
with $\varphi$ as the inflaton. The scenario of a vector field as a curvaton
was discussed in detail in \cite{Di1,Di2}; in such scenario  the vector field does
not drive inflation (to avoid large scale anisotropy) but becomes important after
inflation when it may dominate the Universe and imprint its perturbation spectrum.
A scale invariant spectrum can be generated by $\hat\w_\mu$  provided that during inflation the mass-squared
of $\hat\w_\mu$ is {\it negative} and large in absolute value ($\sim\! H^2$) while after inflation
 is positive and the vector field engages in oscillations and
behaves as pressureless matter; this means
it does not lead to large-scale anisotropy when it dominates \cite{Di1,Di2}.
This scenario cannot apply  here 
since $m^2_\w$ is always positive.
Indeed, the second term in (\ref{twp})
%\medskip
\bea\label{gfi}
\Delta L=\frac12 \sqrt{\hat g} f(\varphi)\, \hat\w_\mu \hat\w^\mu,
\qquad
f(\varphi)
=6\gamma M^2 \cosh^2\frac{\varphi}{2M\sqrt{\gamma}}.
\eea
has $f(\varphi)>0$, for any value of $\varphi$ and the effective mass-squared of $\hat\w_\mu$
is always positive.

Finally, in  Friedmann-Robertson-Walker (FRW) universe 
$\hat g_{\mu\nu}\!=\!(1,-a(t)^2,-a(t)^2,-a(t)^2$,  the
vector field background compatible with the metric is $\hat\w_\mu(t)\!=\!(\hat\w_0(t),0,0,0)$.
However, from the equation of motion of $\hat\w_\mu$  one immediately sees  that
$\hat\w_\mu(t)\!=\!0$, (see also eq.(\ref{oui}) for details).
In this case $\Delta L$ 
is vanishing.
 Therefore, we are left  with potential (\ref{scalars2}) and the usual 
 formalism  of single-field inflation in  Einstein gravity applies, with $\varphi$ as  inflaton.

One may ask what happens at the perturbations level?
One easily sees that
perturbations $\delta\varphi$ of $\varphi$  do not mix with perturbations $\delta\hat\w_\mu$
(of longitudinal mode/Stueckelberg field $\rho$)  of massive $\hat\w_\mu$. Such
mixing is in principle  possible, with potential impact on inflation predictions,
but it vanishes  since it is proportional to $\hat\w_\mu(t)(=0)$, as seen from expanding
$\Delta L$ to quadratic level in perturbations:
$\Delta L\propto \hat\w^\mu(t) \,\delta\varphi\,\delta \hat\w_\mu+\cdots$ \footnote{
The absence of such mixing is also due to 
 the FRW metric and to the fact that  $\rho$ (radial direction)
 and  $\varphi\sim \hat\phi_1$ were orthogonal directions in field space (that do not mix) and
 similar for their perturbations.}.
As a result,  the coupling $\Delta L$  does not affect $\delta\varphi$ and the
predictions of inflation by $\varphi$. For further discussion
on  perturbations $\delta\varphi$ and $\delta\hat\w_\mu$
see Appendix~\ref{ApC} which supports these results.

The above arguments justify our use below of  single-field slow-roll 
formulae\footnote{
With $M\sim\langle\rho\rangle$  a simple phase transition scale, 
values of the field  $\varphi\geq M$ are natural.}
\medskip
\bea\label{eqeps}
\epsilon&=&
\frac{M^2}{2} 
 \Big\{
\frac{V^\prime(\h)}{V(\h)}
\Big\}^2
= 
\frac{{4}}{3}\,\gamma \,\xi_1^2\, \sinh^2 \frac{\h}{M\sqrt {6\gamma}} +\cO(\xi_1^3),\,
\\[4pt]
\eta&=&
M^2\,\frac{V^{\prime\prime}(\h)}{V(\h)}
=
-\frac23\,\xi_1 \,\cosh {\frac{\h}{M\sqrt {6\gamma}}}
+\frac83 \gamma\,\xi_1^2
 \sinh^2 \frac{\h}{2 M\sqrt {6\gamma}}
+\cO(\xi_1^3),\,
\label{eqeta}
\eea

\medskip\noindent
The number of e-folds  is
\medskip
\bea\label{nnn}
N= \frac{1}{M^2} \int_{\varphi_e}^{\varphi_*}  d\varphi 
\,\,\frac{V(\varphi)}{V^\prime(\varphi)}
=
\Big\{-\frac{3}{4\, \xi_1}\ln\tanh^2\frac{\varphi}{2 M \sqrt{6\gamma}}
+ 3\gamma \ln\cosh^2\frac{\varphi}{2 M \sqrt{6\gamma}}\Big\}
\Big\vert_{\varphi=\varphi_e}^{\varphi=\varphi_*}.
\eea

\medskip\noindent
with the last step in (\ref{eqeps}),  (\ref{eqeta}),  (\ref{nnn})
 valid  in the leading approximation $\lambda_1\xi_0\ll \xi_1^2$ needed
for a deep enough minimum for  inflation;  $\varphi_e$ is determined
by $\epsilon(\varphi_e)=1$ and $\h_*$ is the initial value of the scalar field.
Further,  the scalar spectral index
\medskip
\bea\label{nsns}
n_s\, =\, 1 + 2\,\eta_* - 6\, \epsilon_*=
1 
-
\frac43\,\xi_1 \,\cosh\frac{\h_*}{M\sqrt {6\gamma}}
+ 
\frac{8}{3}\,\xi_1^2\,\gamma\,\Big[\cosh^2 \frac{\h_*}{M\sqrt {6\gamma}} -1 \Big]
+
\cO(\xi_1^3).\,
\eea

\medskip\noindent
With the tensor-to-scalar ratio $r=16 \epsilon_*$,  then from (\ref{eqeps}), (\ref{eqeta}), (\ref{nsns})
%\medskip
\bea\label{rns}
r={12}\,\gamma\, (1- n_s)^2 -\frac{64\gamma}{3}\, \xi_1^2+\cO(\xi_1^3).
\eea

\medskip\noindent
The  non-minimal coupling is reducing $r$,  for fixed $n_s$. 
If we ignore  the term $\propto \xi_1^2$  and higher orders, then the Palatini case
 ($\gamma=1$) has a larger $r$ than  Weyl theory ($\gamma=1/4$), for the same $n_s$.
This  is confirmed by exact numerical results, see later.
From (\ref{nnn}), we also find
%\medskip
\bea\label{id}
r\approx \frac{48\gamma}{\overline N^2}+ \frac{64\gamma}{\overline N}\times O(\xi_1);\qquad
n_s\approx 1-\frac{2}{\overline N}+\cO(\xi_1)
\eea
%\medskip\noindent
with $\overline N\approx N+9$ and $\gamma=1/4$
in the Weyl case and $\overline N\approx N+28$ and $\gamma=1$ for the Palatini case. 
Eqs.(\ref{id}) are only an approximation and ignore some $\xi_1$ dependence in $\overline N$,
 but  give an idea of the exact behaviour (see later, Figure~\ref{fig02}).

There is an additional constraint on the parameters
space of Weyl/Palatini models, from the normalization of the CMB anisotropy
 $V_0/(24 \pi^2 M^4 \epsilon_*)=\kappa_0$, $\kappa_0=2.1\times 10^{-9}$
and with $r<0.07$ \cite{planck2018} then 
$\xi_0=1/(\pi^2 r \kappa) \geq 6.89 \times 10^8$. With this bound, 
condition $\lambda_1 \xi_0\!\ll \!\xi_1^2$ 
 is respected for any perturbative $\xi_1$, $1/\xi_0$,
by choosing an ultraweak $\lambda_1\!\ll\! \xi_1^2/\xi_0$.

Let us compare eq.(\ref{rns}) to that in the Starobinsky model
of
\bea
\cL= (-1/2) M^2\, R+ (\xi_0/4!)\, R^2,
\eea
giving
$V\!=\!V_0 \big(1-e^{-\varphi\sqrt{2/3}/M}\, \big)^2$ with
$V_0\!=\!3 M^4/(2\xi_0)$; then  $r\!\approx\! 12/N^2$,  $n_s\!\approx\! 1 -2/N$ 
and 
%\medskip
\bea\label{sssta}
r= 3\, (1-n_s)^2.
\eea
%\medskip
Interestingly, in eq.(\ref{rns}) with  $\xi_1\!\sim\! 10^{-3}$ 
or smaller, the term $\propto\xi_1^2$ and higher powers
have a negligible correction to  $r$ and $(1-n_s)^2$ and
can be ignored; therefore  Weyl inflation  ($\gamma=1/4$) 
recovers   relation (\ref{sssta})  of  Starobinsky model \cite{Sta,planck2018}.
For larger values of $\xi_1$ and fixed $n_s$,  $\xi_1$ reduces $r$ of Weyl inflation 
below that of Starobinsky model.
In the Palatini case  relation (\ref{sssta})  is not possible
 (unless $\xi_1$ is tuned  for every  $n_s$) -
the slope of  $r(n_s)$ is different.

\subsection{Numerical results}

\begin{wrapfigure}{R}{0.43\textwidth}
\vspace{-0.5cm}
\centering
\includegraphics[height=0.39\textwidth]{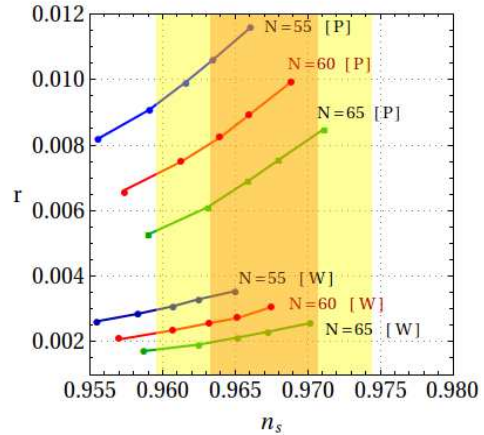}
\medskip
\caption{\small
The dependence $r(n_s)$ for various curves of constant $N$  (as shown),
 for the Palatini theory ([P])  and Weyl theory ([W]). The dots 
stand for  $\xi_1$ values  corresponding to the curves in the last two plots 
in Figure~\ref{fig01}.
The yellow (orange) area corresponds to the 
range  measured for $n_s$ at $95 \%$CL $(68 \%)$, respectively.
The curves show the largest range of values for $r$ 
with $\lambda_1\xi_0\ll  \xi_1^2$; this range  shrinks towards
smaller $r$ if $\lambda_1\xi_0$  increases to $\lambda_1\xi_0\sim \xi_1^2$. }
\label{fig02}\vspace{-0.2cm}
\end{wrapfigure}

Our exact numerical results (with no expansion in powers of $\xi_1$)
 are given by the plots of potential, $(n_s,N)$,  $(n_s,r)$,
presented in figures~\ref{fig01} and~\ref{fig02} for Weyl and Palatini cases.
Their differences are due to different  $\gamma$.
The results show a value of $r$  smaller in the Weyl case than in Palatini case,
for relevant $n_s$. For  $n_s=0.9670\pm 0.0037\, (68\% {\rm CL})$ 
(TT, TE, EE + low E + lensing + BK14 +BAO)  \cite{planck2018} one finds
%\medskip
\bea\label{tre1}
\!\!\!\!{\rm Palatini:} &&\!\!\!\!\!  N=60,\quad   0.00794\leq r\leq 0.01002,\quad\\
{\rm Weyl:} && \!\!\!\!\! N=60,\quad     0.00257\leq r\leq 0.00303.
\eea
and for $n_s$ at $95\% {\rm CL}$ one has
%\medskip
\bea
\!\!\!\!{\rm Palatini:} &&\!\!\! \!\!  N=60,\quad 0.00700\leq r\leq 0.01002,\quad \\
{\rm Weyl:} &&\!\!\!\!\! N=60,\quad  0.00227\leq r\leq 0.00303.\label{tre2}
\eea

The case of Starobinsky model for $N=60$ corresponds to the upper limit of $r$ (0.003) of
 the Weyl model (top curve in figure~\ref{fig01} and highest $r$ in figure~\ref{fig02} for $N=60$),
 while in the Palatini case a larger $r$ is allowed for the same $n_s$, $N$.

While the plots  in figure~\ref{fig01} have $\lambda_1\xi_0=10^{-8}$, 
they are actually more general.
In the extreme case of  $\lambda_1\!\sim\! 0$, 
corresponding to a simplified potential (without the last term in (\ref{scalars2})),
 the same range of values for $(n_s,r)$  shown in this figure remains
 valid.
 However, if we increase $\lambda_1\xi_0$ to  $\lambda_1\xi_0 \approx \xi_1^2$, 
the last term in (\ref{scalars2}) becomes relatively large, 
 the rightmost curves of $V$  (of smallest $\xi_1$)  have their minimum lifted   
and the range  for  $(n_s,r)$  in (\ref{tre1}) to (\ref{tre2}) is reduced:
the  smaller values  $\xi_1\!\sim\! 10^{-3}$ in  Figure~\ref{fig01},
cannot then have successful inflation. 
 
{
The main results of this work are summarised in figure~\ref{fig02}; in this figure
the dependence $r(n_s)$ is shown for different curves of constant $N$, 
that respect the required parametric constraint $\lambda_1\xi_0\leq \xi_1^2$.
The curves $r(n_s)$ give a numerically exact representation of the dependence in
eq.(\ref{rns}); they are  extended even outside the $95\%$ CL range for $n_s$. 
In all cases,  the Palatini case  has $r$  larger than in the Weyl
case. This aspect and the different slope of the curves $r(n_s)$
can be used to distinguish these models from each other and from other
models in  future experiments. }

The small   $r$  predicted by both Weyl and Palatini gravity models may be
reached  by the next generation of CMB experiments: 
CMB-S4, LiteBIRD, PICO, PIXIE  \cite{CMB1,CMB4,CMB3,litebird,Pixie,CMB2}
that will reach a precision for $r$ of   $\sim 5\times 10^{-4}$.
Therefore they will be able to test these two  
inflation models.

\subsection{Corrections and other models}

Compared  to another   model  with Weyl gauge symmetry \cite{ghilen} (figure 2)
which is {\it linear} in $R(\tGamma,g)$ and had $r\sim 0.04 - 0.06$, we see that
the presence of the $R^2(\tGamma,g)$ term in the Weyl theory  reduced $r$ significantly
(for a fixed $n_s$).  Reducing $r$ by an $R^2$ term also exists in the 
Palatini models without Weyl gauge symmetry \cite{Palatini2}.
 Therefore, a small measured $r\!\sim\! 10^{-3}$ 
may indicate a preference for  quadratic gravity models of inflation.

The Weyl inflation case of $\xi_1= 10^{-3}$ or smaller 
is similar to the Starobinsky model. % $(-1/2) M^2 R+\xi_0 R^2$.
Here we have  an additional scalar field\footnote{This may be the Higgs, see discussion in the
  text after eq.(\ref{fghj}).}, with $\varphi$  playing the role of the inflaton\footnote{
To be exact inflation is mostly due to $\phi_0$, first term in (\ref{scalars2}),(\ref{toE}),
  hence the similarity to Starobinsky case.}.
The other scalar in the  Weyl theory (radial direction $\rho$ in
the $\phi_0$, $\phi_1$ space) is used to generate  the Planck scale and the mass of $\w_\mu$.
Briefly,  Weyl gravity gives a relation $r(n_s)$ similar to that in the Starobinsky model,
with similar, large $\xi_0$, while also providing protection against corrections to $r$ from
higher dimensional operators; these  are  forbidden since their effective scale
violates the symmetry; the Stueckelberg field cannot play the role
of this scale since it was eaten by the Weyl  field to all orders.
Another benefit for Weyl inflation is the minimal approach: one
only needs to consider the SM Higgs field in the Weyl conformal geometry;  the underlying
geometry provides the spontaneous breaking of the  Weyl quadratic gravity
action to  the Einstein action and the Planck scale generation.

Despite this similarity of the Weyl and the Starobinsky models, it is possible
to distinguish between them; it may happen that a curve $r(n_s)$ corresponding
to $\xi_1>10^{-3}$ is preferred by data (see $r(n_s)$ curves in figure~\ref{fig01}), in which case
it is shifted below that of the Starobinsky model for the same $n_s$
- the two models are  distinguishable.
Also the Weyl model has an additional  coupling, see  $\Delta L$ in (\ref{gfi}).
While $\Delta L$ does not mix linear perturbations of $\delta \w_\mu$ and of
$\delta\varphi$, it can lead however to cubic interactions of the
form $f'(\varphi) \delta\varphi\delta \w_\mu\delta\w^\mu$.  These can
result in different predictions for the inflationary bispectrum compared
to the pure single-field case. This can be used to further distinguish the Weyl
case  from the  Starobinsky $R^2$ inflation (for $\xi_1\sim 10^{-3}$).
The analysis of non-Gaussianity is  thus interesting for
further research.

{The above results are subject to corrections from other  operators of $d=4$  that
may exist and are Weyl gauge invariant, as we discuss below.}

In the Weyl case the Weyl-tensor-squared operator 
of Weyl geometry may be present $(1/\zeta) \tilde C_{\mu\nu\rho\sigma}^2$.
This can be re-written in a metric description as the  Weyl-tensor-squared term of 
Riemannian geometry $(1/\zeta)\, C_{\mu\nu\rho\sigma}^2$ plus a gauge kinetic term 
of $\w_\mu$ which gives a threshold correction to our coupling $\alpha$. 
The Weyl tensor term is invariant under Weyl gauge transformations performed to reach 
the Einstein-Proca action, hence one simply adds it to the final action, eq.(\ref{W3}).
This operator has an impact on the value of $r$ that we found numerically and in
eq.(\ref{id}) with $\gamma=1/4$ for Weyl case.
The  overall impact of the Weyl tensor term is essentially a  rescaling of $r$
into\footnote{{For an extended analysis  of the role of the Weyl tensor term
    on inflation (in particular in $R^2$ inflation)
    see \cite{Baumann,Manheim1,Star1}.
    The above mentioned rescaling effect of a Weyl tensor squared term on the value of $r$ found in
    its absence (e.g. Starobinsky result) is more general; for example, for a
    non-local   Weyl-tensor-squared  term (of Riemannian geometry), the effect is again a
  rescaling of $r$ value found in its absence, albeit  by an overall 
  factor different from that above \cite{Star1}; the different factor is
  due to the more general structure of this term.  Such Weyl tensor-dependent
  operator cannot appear here  since it is forbidden by the Weyl gauge symmetry. }}
$r_c=r\,(1+ 8/(\zeta\,\xi_0))^{1/2}$ \cite{Winflation}.
Since our $\xi_0$ is large, only a  low $\vert\zeta\vert\sim 1/\xi_0$ can increase $r$
and this comes with an instability since the mass of the associated spin-two ghost (or tachyonic)
state that this operator brings  is $m^2\sim \zeta\, M^2$, where $M$ is the Planck scale. 
Therefore, a stable Weyl gravity 
model up to the Planck scale will not modify the value of $r$. Other operators
in Weyl gravity are topological and do not affect $r$ (classically).

In the Palatini case one should consider the remaining quadratic  operators of $d=4$ \cite{Borunda}
that are  Weyl gauge invariant and have a symmetric connection.
They modify the equation of motion of $\tGamma$  and
the vectorial non-metricity (\ref{nonm}); unfortunately, it does not seem possible to find
in this case an analytical solution to this equation due to its modified,
complex structure and new states
present (ghosts, etc). Additional simplifying assumptions would be needed, making
the analysis model dependent. We only mention here
the interesting possibility that for a symmetric $\tGamma$, the solution 
$\tGamma$ may become equal to that  in  Weyl-geometry (\ref{o});  if so, the Palatini
approach would provide an ``offshell'' version of Weyl quadratic gravity that is 
recovered for  $\tGamma$ onshell.

\section{Conclusions}

We made a comparative study  of the action and inflation in two theories of quadratic gravity
 with Weyl gauge symmetry: the original  Weyl gravity action 
and the  Palatini version of the same  action, obtained by replacing the Weyl connection 
by  Palatini connection. The actions of these theories are non-minimally coupled to a 
(Higgs-like) field $\phi_1$. 

Given the symmetry, there  is no scale in these theories.
Mass  scales are generated by an elegant spontaneous breaking of gauged scale symmetry that happens
even {\it in the absence of matter}: the necessary scalar field (Stueckelberg field $\phi_0$) is
not added {\it ad-hoc} as usually done to this purpose, but is of geometric origin and is
''extracted''
from the $R(\tGamma,g)^2$ term in the action. If matter ($\phi_1$)
is present, the Stueckelberg field is actually the radial direction ($\rho$) in the field space
of $\phi_0$ and $\phi_1$; the  field $\rho$ is then eaten by the Weyl gauge field $\w_\mu$ which
acquires mass $m_\w\sim\langle\rho\rangle$ near the Planck scale.
 The breaking conserves  the number of degrees of freedom
 and generates in the broken phase the Einstein-Proca action  for  $\w_\mu$.
In both theories, below the mass of $\w_\mu$
the connection becomes Levi-Civita and Einstein gravity is recovered,
with  an ``emergent''  Planck scale  $M\sim \langle\rho\rangle$
and a scalar potential (of the remaining, angular-variable field $\varphi$).

The  potential $V(\varphi)$ is controlled by the symmetry of the theory 
together with effects from the  non-trivial  connection $\tGamma$,  different in the two theories.
For {\it  small} field values, $V$  is similar in both theories;
the scalar field can act as  the Higgs field, in which case
the potential displays  electroweak symmetry  breaking. 
For {\it large}  field values, the potential has the same  form 
in Weyl and Palatini theories up to couplings and field rescaling (due to 
different non-metricity) and gives successful inflation. 

Our main results, comparing inflation predictions
 in the two theories and summarised in Figure~\ref{fig02},
showed how a different non-metricity impacts on inflation predictions.
In Weyl gravity the scalar-to-tensor ratio $0.00257\!\leq \!r\!\leq\! 0.00303$,
which is smaller than in Palatini case, $0.00794\! \leq\! r\!\leq\! 0.01002$,
 for measured $n_s$ at  68$\%$ CL and $N=60$ e-folds. 
Similar results exist for $n_s$ at $95\%$CL or  mildly different $N$, etc.
Such values of $r$ will be measured by new CMB experiments that 
can then test  and distinguish Weyl and Palatini quadratic gravity.

There are  similarities of inflation in  Weyl and Palatini cases to Starobinsky inflation ($R+\xi_0 R^2$).
In  Weyl and Palatini theories one also has an $R^2$ term with a large   $\xi_0$ that reduces $r$,
but  there is also a  non-minimally coupled scalar field ($\phi_1$);
one combination of fields  is acting as the  inflaton while
the  other  (radial)  combination enabled the  breaking of the
gauged scale  symmetry and the generation of mass scales (Planck, $\w_\mu$ mass).
In both Weyl and Palatini theory, for a fixed $n_s$,  reducing  the 
non-minimal coupling ($\xi_1$) increases $r$ 
which in Weyl theory is bounded from above by that of  Starobinsky inflation.
Unlike in the Palatini theory, Weyl  gravity  for $\xi_1\leq 10^{-3}$  gives
a dependence $r(n_s)$  essentially similar to that in Starobinsky inflation,
while also  protecting  $r$ against    higher dimensional operators corrections. 

\bigskip\bigskip\bigskip

\begin{center}
---------------------------
\end{center}

\vspace{0.9cm}

\section*{Appendix}

\bigskip

\def\theequation{A-\arabic{equation}}
\def\thesubsection{A}
\setcounter{equation}{0}
\def\thefigure{A-\arabic{figure}}

\subsection{Weyl gravity}\label{ApA}

We include here  basic  information on Weyl gravity used in the text.
First, in the (pseudo)-Riemannian case (Einstein gravity) $\nabla_\mu g_{\alpha\beta}\!=\!0$ with
$\nabla_\mu$ defined by the  Levi-Civita connection
\medskip
\bea\label{eom1}
\Gamma_{\mu\nu}^\rho (g)= 
(1/2)\, g^{\rho\beta}\,(\partial_\nu g_{\beta\mu}+\partial_\mu g_{\beta\nu}-\partial_\beta g_{\mu\nu}).
\eea

\medskip\noindent
Setting $\nu=\rho$ and summing over gives $\Gamma_\mu\equiv \Gamma_{\mu\nu}^\nu=\partial_\mu \ln \sqg$ used in 
the text.

In Weyl gravity and  conformal geometry the theory has vectorial  non-metricity, i.e.
\medskip
\bea\label{gg}
\tilde\nabla_\lambda\, g_{\mu\nu}=-\w_\lambda \, g_{\mu\nu},
\eea

\medskip\noindent
so $\w_\lambda=(-1/4)\, g^{\mu\nu}\,\tilde\nabla_\lambda g_{\mu\nu}$; here $\tilde\nabla_\mu$ is
defined by the Weyl connection  $\tilde \Gamma_{\mu\nu}^\rho$:
\medskip
\bea\label{nvv}
\tilde\nabla_\lambda g_{\mu\nu}=
\partial_\lambda g_{\mu\nu} - \tGamma_{\mu\lambda}^\rho g_{\rho\nu}-\tGamma_{\nu\lambda}^\rho g_{\mu\rho}.
\eea

\medskip\noindent
Write this for cyclic permutations of the indices and combine the three equations to
find 
\medskip
\bea
\tGamma_{\mu\nu}^\rho=\Gamma_{\mu\nu}^\rho(g)+
(1/2)\, g^{\rho\lambda}\,(\tilde\nabla_\lambda g_{\mu\nu}-\tilde\nabla_\mu g_{\nu\lambda}
 - \tilde\nabla_\nu g_{\lambda\mu}),
\eea

\medskip\noindent
which with (\ref{gg}) gives the  Weyl connection
\medskip
\bea\label{o}
\tilde\Gamma_{\mu\nu}^\rho=
\Gamma_{\mu\nu}^\rho(g)+(1/2)\,\big[\delta_\mu^\rho\,\w_\nu+\delta_\nu^\rho\,\w_\mu-g_{\mu\nu}\,\w^\rho\big].
\eea

\medskip\noindent
  $\tilde \Gamma_{\mu\nu}^\rho$ are symmetric ($\tilde \Gamma_{\mu\nu}^\rho=\tilde \Gamma_{\nu\mu}^\rho$)
i.e. there is no torsion. 
$\tGamma$ is invariant under transformations (\ref{cr}), (\ref{s2})
since the  variation of the metric is compensated by that of  $\w_\mu$. 
Using that
$g^{\alpha\beta}\tilde\nabla_\lambda g_{\alpha\beta}=2\tilde\nabla_\lambda\ln \sqg$ one 
finds for the Weyl field
\medskip
\bea\label{www}
\w_\lambda=(-1/2)\,\,\tilde\nabla_\lambda \ln\sqg.
\eea

\medskip\noindent
Setting $\nu=\rho$  in (\ref{o}) and summing over, we recover our definition (\ref{s2})
in the text:
\medskip
\bea
\tGamma_\mu=\Gamma_\mu(g)+2\, \w_\mu.
\eea 

\medskip The Riemann and Ricci  tensors  in Weyl geometry are defined as 
in  Riemannian geometry but with the replacement of the Levi-Civita connection $\Gamma_{\mu\nu}^\rho(g)$
 by the
new $\tilde \Gamma_{\mu\nu}^\rho$
\medskip
\bea
 R^\lambda_{\,\mu\nu\sigma}(\tGamma,g)=
 \partial_\nu \tilde\Gamma^\lambda_{\mu\sigma}
 -\partial_\sigma\tilde\Gamma^\lambda_{\mu\nu}
 + \tilde\Gamma^\lambda_{\nu\rho}\,\tilde\Gamma^\rho_{\mu\sigma}
 -\tilde\Gamma^\lambda_{\sigma\rho}\,\tilde\Gamma^\rho_{\mu\nu},
\eea
 and 
\bea
R_{\mu\sigma}(\tGamma,g)=R^\lambda_{\,\,\mu\lambda\sigma}(\tGamma,g),\qquad
R(\tGamma,g)=g^{\mu\sigma}\,\cR_{\mu\sigma}(\tGamma,g).
\eea

\medskip\noindent
 Since $\tGamma$ is invariant under 
transformations  (\ref{cr}), (\ref{s2}), then the Riemann and Ricci tensors of Weyl 
geometry are also invariant. Since the Weyl scalar curvature $R(\tGamma,g)$ contains $g^{\mu\nu}$, 
it transforms covariantly
\medskip
\bea\label{RR}
\hat{R}(\tGamma,g)= (1/\Omega^2)\,R(\tGamma,g).
\eea

\medskip\noindent
This helps build Weyl gauge  invariant  operators.
Using the expression of $\tGamma$,  one shows 
\medskip
\bea\label{Wcurvature}
R(\tGamma,g) =
R(g) - 3 \,   \nabla_\mu \w^\mu -\frac32 \, g^{\mu\nu}\,\w_\mu \w_\nu,
\eea

\medskip\noindent
where $R(g)$ is the  Riemannian scalar curvature
and $\nabla_\mu\w^\mu$ is defined by Levi-Civita connection.
Eq.(\ref{Wcurvature}) was used in the text, in going from (\ref{ll1prime}) to (\ref{ll2})
for the Weyl case.

\def\theequation{B-\arabic{equation}}
\def\thesubsection{B}
\setcounter{equation}{0}
\def\thefigure{B-\arabic{figure}}

\subsection{Palatini gravity}\label{ApB}

We present here the connection and the scalar curvature for the Palatini 
approach to gravity, used in the text.
In this case,   similarly to Weyl gravity,  $\tGamma$  is 
not determined by the metric (apriori is independent of it), hence it is invariant under
rescaling $g_{\mu\nu}$. The connection is determined by its equation of motion from the 
 Lagrangian of eq.(\ref{ll1prime}).  Solving this equation of motion  one finds 
\cite{gPalatini} (eqs.23, 25, 39)
\medskip
\bea\label{nabla2}
\tilde\nabla_\lambda (\rho^2 g_{\mu\nu})
=(-2)\rho^2 (g_{\mu\nu}\,V_\lambda-g_{\mu\lambda} V_\nu - g_{\nu\lambda} V_\mu),
\eea

\medskip\noindent
where $V_\lambda$ is some arbitrary vector, related to $\w_\lambda$ (see below).
One writes (\ref{nabla2}) for cyclic permutations of the three indices, then combines the equations  
obtained and uses the symmetry $\tGamma_{\mu\nu}^\alpha=\tGamma_{\nu\mu}^\alpha$, to find
\medskip
\bea\label{Vm1}
\tGamma_{\mu\nu}^\alpha&=&\Gamma_{\mu\nu}^\alpha(\rho^2 g)
-\,\big(3\, g_{\mu\nu}\, V_\lambda
-g_{\nu\lambda}\,V_\mu - g_{\lambda\mu} \,V_\nu\,\big)\,g^{\lambda\alpha},
\eea
where
\bea\label{utr}
\Gamma^\alpha_{\mu\nu}(\rho^2 g)=\Gamma^\alpha_{\mu\nu}(g)+1/2\, \big( \delta_\nu^\alpha
\, \partial_\mu +\delta_\mu^\alpha\, \partial_\nu -g^{\alpha\lambda} g_{\mu\nu} \,\partial_\lambda)
\ln\rho^2,
\eea

\medskip\noindent
with $\Gamma_{\mu\nu}^\alpha(g)$ the Levi-Civita connection for $g_{\mu\nu}$. 
Setting $\nu\!=\!\alpha$  in (\ref{Vm1}) one then finds 
$\tGamma_\mu\!=\Gamma_\mu(\phi^2 g)+ 2\,  V_\mu$ 
and from (\ref{utr}): $\Gamma_\mu(\phi^2 g)=\Gamma_\mu(g)+2 \,(\partial_\mu\ln\rho^2)$.
From these two equations and with  the definition $\w_\lambda = 1/2\, (\tGamma_\mu -\Gamma_\mu(g))$, 
then $V_\lambda=\w_\lambda-\partial_\lambda\ln\rho^2$. 
Using this relation  and that found  by contracting (\ref{nabla2}) by $g^{\mu\nu}$, then 
\medskip
\bea
\w_\lambda=(-1/2) \tilde\nabla_\lambda \ln\sqrt{g}.
\eea

\medskip\noindent
similar to (\ref{www}), but with different $\tGamma$. 
Finally, eqs.(\ref{Vm1}), (\ref{utr}) together with 
$V_\lambda=\w_\lambda-\partial_\lambda\ln\rho^2$, give the expression of 
 $\tGamma$ in terms of $g_{\mu\nu}$, $\rho$ and $\w_\lambda$ 
and verifies  that $\tGamma$ is indeed invariant under a gauged scale transformation 
(\ref{cr}), (\ref{s2}). This is obvious  since $\rho^2 g_{\mu\nu}$ and $V_\mu$
are invariant in  (\ref{Vm1}).  With $\tGamma$ a  function of $\w_\lambda$, $\phi$, $g_{\mu\nu}$,
one computes  the Ricci tensor $R_{\mu\nu}(\tGamma)$ for Palatini gravity,
 then the scalar curvature $R(\tGamma,g)=g^{\mu\nu} R_{\mu\nu} (\tGamma,g)$.
The result is \cite{gPalatini}:
\medskip
\bea\label{Pcurvature}
R(\tGamma,g)=R(g)-6 g^{\mu\nu} \nabla_\mu\nabla_\nu\ln\rho
-6 (\nabla_\mu\ln\rho)^2
-12\, \big(\nabla_\lambda V^\lambda
+ V^\lambda \partial_\lambda\ln\rho^{{2}}\big)
-6  V_\mu\, V^\mu,
\eea

\medskip\noindent
with $R(g)$ the  Ricci scalar (Riemannian case),  and 
 $V_\lambda\equiv w_\lambda-\partial_\lambda\ln\rho^2$. Replacing (\ref{Pcurvature})
in eq.(\ref{ll1prime}) for the Palatini case, one finds 
after some algebra eq.(\ref{ll2}) in the text with $\gamma=1$.
At the same time, the vectorial non-metricity becomes
\medskip
\bea\label{nonm}
\tilde\nabla_\lambda g_{\mu\nu}=(-2)\,(g_{\mu\nu} \w_\lambda-g_{\mu\lambda} \w_\nu-g_{\nu\lambda} \w_\mu),
\eea

\medskip\noindent
which is different from (\ref{gg})  of Weyl geometry,
but has the same trace $g^{\mu\nu}\tilde\nabla_\lambda g_{\mu\nu}=-\w_\lambda/4$.

\vspace{1cm}

\def\theequation{C-\arabic{equation}}
\def\thesubsection{C}
\setcounter{equation}{0}
\def\thefigure{C-\arabic{figure}}

\subsection{Inflation:  perturbations to the scalar and vector fields}
\label{ApC}

We discuss in detail the scalar ($\delta\phi$) and vector ($\delta\w_\mu$)
fields perturbations in a FRW universe  $g_{\mu\nu}\!=\!(1, -a(t)^2,  -a(t)^2,  -a(t)^2)$
and show that  $\Delta L$ of (\ref{gfi})  does not affect  inflation by $\varphi$.
To simplify notation hereafter we remove the 'hat' ($\hat\,$) on $\w_\mu$,  $g_{\mu\nu}$
when we refer to action (\ref{twp}).

\bigskip\noindent
$\bullet$ Let us first review the usual
case of a single scalar field, see e.g.\cite{Riotto}, needed later.
Consider
\bea\label{phikt}
\cL_\varphi=\int \sqrt{g} \Big[\frac12 \,g^{\mu\nu} \,\partial_\mu\varphi\partial_\nu\varphi
-\mathcal{V}(\varphi)\Big].
\eea
The equation of motion   $\nabla_\mu\nabla^\mu \varphi + \mathcal{V}^\prime(\varphi)=0$ gives
for a FRW metric:
\bea\label{back}
\ddot \varphi(\vec x,t)+ 3 H \dot\varphi(\vec x,t)-\frac{1}{a(t)^2} \partial_j\partial_j
\varphi(\vec x,t)+\cV^\prime(\varphi)=0
\eea
Expanding about $\varphi(t)$, with
$\varphi(\vec x,t)=\varphi(t)+\delta\varphi(\vec x,t)$, one has
at linear level 
\bea
\delta\ddot \varphi(\vec x,t)+ 3 H \delta\dot\varphi(\vec x,t)-\frac{1}{a(t)^2} \partial_j\partial_j
\delta\varphi(\vec x,t)+\cV''(\varphi(t))\delta\varphi(\vec x,t)=0
\eea
Using mode expansion
$\delta\varphi(\vec x,t)=\int d^3k/(2\pi)^{3/2} \delta\varphi_k(t) \exp( i \vec x \vec k)$, then
\medskip
\bea\label{j3}
\delta\ddot \varphi_k+ 3 H \delta\dot\varphi_k+
\big[ k^2/a^2+\cV''(\varphi(t))\big] \delta\varphi_k=0
\eea
or, with a notation $\delta\phi_k=\delta\chi_k/a(t)$
\medskip
\bea
\delta\ddot \chi_k+  H \delta\dot\chi_k+
\big[ k^2/a^2-\dot H -2 H^2+\cV''(\varphi(t))\big] \delta\varphi_k=0.
\eea

\medskip\noindent
In conformal time  ($\eta$) via $dt=a(t)^2 d\eta$, this equation  becomes
\medskip
\bea\label{h3}
\delta \chi''_k+ 
\big[ k^2 -(1/\eta^2)\, (\nu^2-1/4)\big] \delta\varphi_k=0, \qquad
\nu^2=9/4-\cV''/H^2. 
\eea

\medskip\noindent
where we used that with $a\sim e^{H t}$ then
$a(\eta)=-1/(H\eta)$ and $2/\eta^2=a''/a$  with $H\sim$constant.
In the subhorizon limit $-\eta\, k\gg  1$ the solution should be
$\delta\chi_k=e^{-i k \eta}/\sqrt{2 k}$. With this boundary condition,
the solution is
%\medskip
\bea\label{sol2}
\delta\chi_k=\frac{\sqrt{\pi}}{2} e^{i (\nu+1/2) \pi/2}\sqrt{-\eta}\,H_\nu^{(1)}(-\eta k)
\eea
%\medskip\noindent
where $H^{(1)}$ is the Hankel function of first kind.
This  leads to the usual power spectrum, with
\bea\label{hhh}
\vert\delta\phi_k\vert^2\approx \frac{H^2}{2 k^3}\Big(\frac{k}{a H}\Big)^{2\eta_\phi}, \qquad
P_{\delta\phi_k}
=\frac{k^3}{2\pi^2} \vert\delta\phi_k\vert^2=\Big(\frac{H}{2\pi}\Big)^2\,\Big(\frac{k}{a H}\Big)^{2\eta_\phi}
\eea
with $\eta_\phi=3/2-\nu\approx \cV''/(3 H^2)=M_p^2\,\cV''/\cV\ll 1$. This gives $n_\phi=1+2\eta_\phi$
($H\sim$ constant).

For later use, we also consider solution (\ref{sol2})
when  $\cV''(\varphi)\!>\! (9/4)\, H^2$
i.e. $\nu$ is imaginary, $\nu=i\tilde\nu$, $\tilde\nu$  real.
In the (superhorizon) limit $(-\eta\, k)\ll 1$ one finds:
%\medskip
\bea
\delta \chi_k=\frac{(1+i)\, 2^{-3/2-i\tilde\nu} e^{-\tilde\nu\pi/2}}{\sqrt{\pi k} \Gamma(1+i \tilde\nu)}
\Big[\pi (-\eta k)^{2 i \tilde\nu} (1+\coth \pi\tilde\nu)+2^{2i\tilde\nu} \tilde\nu\, \Gamma(i\tilde\nu)^2\Big].
\eea
Returning to  $\delta\phi_k$ notation, one finds (see e.g. \cite{Riotto})
\bea\label{zx3}
P_{\delta\phi_k}
=\frac{\pi e^{-\pi\tilde\nu}}{2} 
\frac{H^2}{(2\pi)^2}
\Big(\frac{k}{a H}\Big)^3\,\{....\}
\eea
where the brackets $\{...\}$ stand for terms that vanish when $\tilde\nu\ra \infty$ 
($\cV''(\varphi)\gg H^2$). Therefore,   modes  $\delta\phi_k$ of $\nu$ imaginary 
are  exponentially suppressed \cite{birrel}; this is expected since they
are too massive  to be excited.

\bigskip

\medskip\noindent
$\bullet$ Consider now our action  (\ref{twp}); its $\varphi$-dependence
is described  by $\cL_\varphi$  by replacing in (\ref{phikt})
\bea\label{11}
\cV(\varphi)\ra \cV(\varphi,\w)=V(\varphi)-\frac12 f(\varphi)\w_\mu w^\mu
\eea
with $V(\varphi)$ of eq.(\ref{scalars2}) and $f(\varphi)$ of eq.(\ref{gfi}).
In this case (\ref{back}) and the equation for $\varphi(t)$
receive a correction from the last term in the rhs
of (\ref{11}). Then  eq.(\ref{j3})
for the perturbations $\delta\varphi_k$, also  with
$\w_\mu(\vec x,t)=\w_\mu(t)+\delta \w_\mu(\vec x,t)$, is now  modified into
%\medskip
\bea\label{j3prime}
\delta\ddot \varphi_k+ 3 H \delta\dot\varphi_k+
\big[ k^2/a^2+\cV''(\varphi(t),\w(t))\big] \delta\varphi_k=f^\prime(\varphi(t)) \w_\mu(t) \delta w^\mu(\vec k,t),
\eea
%\medskip\noindent
where the second derivative $\cV''$
is with respect to $\varphi$ and $\delta \w_\mu (\vec k,t)$ are the Fourier modes of
$\delta w_\mu(\vec x,t)=\int d^3k/(2\pi)^{3/2} \delta \w_\mu(\vec k, t) \exp (i\vec k \vec x)$.
Next, the background $\w_\mu(t)$ compatible with the FRW metric is
$\w_\mu(t)\!=\!(\w_0(t),0,0,0)$, while from (\ref{twp}), 
the equation of motion  of $\w_\mu$ gives
%\vspace{-0.1cm}
\bea\label{oui}
\frac{1}{\sqrt g}\,\partial_\rho\, \big[\sqrt g\, %\nabla_\rho
 F^{\rho\mu}\,\big] + f(\varphi)\, \w^\mu=0.
\eea
One has a trivial solution $w_\mu(t)\!=\!0$  ($f(\varphi)\!\not=\!0$).
Therefore, in (\ref{j3prime}) we must  replace $\cV''(\varphi(t),\w(t))
\ra \cV''(\varphi(t),0)= V''(\varphi)$  while the
rhs of (\ref{j3prime}) is vanishing.
Therefore  equation (\ref{j3prime})  of $\delta\phi_k$ is actually
independent of  $\w_\mu$ and $\delta\w_\mu$ and
there is no mixing of $\delta\varphi$ to $\delta\w_\mu$
\footnote{Apriori a mixing may exist   of   longitudinal
  mode and $\varphi$, and of their perturbations ($\delta\varphi$, $\delta\w_\mu$).}.
Then the calculation of  $\delta\varphi_k$
proceeds as earlier but for  potential $V(\varphi)$, see (\ref{hhh}) for $\cV(\varphi)\!\ra\! V(\varphi)$.
Thus $\Delta L$ does not impact on $\varphi$-inflation and
the usual formulae of single-field inflation in Einstein gravity
apply, as used in Section~\ref{rs3}.

\vspace{0.7cm}
\noindent
$\bullet$
We saw above that the perturbations $\delta\varphi$ do not mix with those
of $\w_\mu$  and $\varphi$-inflation decouples from
$\w_\mu$ in a FRW universe. While  somewhat beyond the purpose of this  work,
  we also examine below  the vector field perturbations,
following \cite{Di1,Di2},  in the approximation $H\sim$ constant. 
Compatibility with the FRW  metric demands computing the perturbations about a
background  $\w_\mu(t)=0$ as seen earlier. In fact 
we may take a more general background,  if initially the vector field
contribution to the stress-energy tensor is negligible relative to that of the scalar,
in  an isotropic universe; we shall then consider  a quasi-homogeneous field
$\partial_i w^\alpha=0$. Our FRW case is always
restored by setting anywhere below $\w_\mu(t)=0$. Then from (\ref{oui})  for
$\mu=0$ and $\mu=i$, respectively
\bea
\w_0(t)=0
\quad \textrm{and}\quad
\ddot \w_i(t)+ H \dot\w_i(t) + f(\varphi) \,\w_i(t)=0.
\eea

\medskip\noindent
In an expanding FRW universe
the relevant physical quantity is not $\w_i$ ($i=1,2,3$) but $q_i=\w_i/a$, as also seen from the
norm $\w_\mu\w^\mu=w_0^2-(w_i w_i)/a^2$ (sum over $i$) and from the stress energy tensor
\cite{Mukh,Di1}.
Then the  last equation becomes
\bea
\ddot q_i+ 3 H \dot q_i+ (2 H^2+ f(\varphi))\, q_i=0.
\eea
Denote $m^2=2 H^2+f(\varphi(t))$ where $f(\varphi)>0$ since
 $f(\varphi(t))=6\gamma M^2\{1+\sinh^2[\varphi(t)/(2 M\sqrt\gamma)]\}$,
 eq.(\ref{gfi}).
Ignoring the time dependent part in $f(\varphi)$, the solution is
\medskip
\bea\label{qqq}
q_i(t)\propto 1/a(t)^{3/2} (c_1\, e^{- \alpha t/2}
+c_2\, e^{\alpha t/2}), \qquad \alpha= \sqrt{H^2-4 m^2}.
\eea

\medskip\noindent
with constants $c_{1,2}$.
Since $\alpha$ is purely imaginary, during inflation the vector
field is massive with damped oscillations (up to corrections due to $\varphi(t))$).
Its contribution to the stress energy tensor ($T^\mu_\nu$)  is 
anisotropic; the spatial part of this tensor contains off-diagonal entries of comparable size to
the diagonal ones and can be made diagonal for a particular  direction of the vector field.
However, with  $q_i(t)\sim 1/a(t)^{3/2}$, the contribution of the vector field to $T^\mu_\nu$
during inflation  is suppressed by the scale factor $1/a(t)^3$
relative to that of\footnote{A diagonal stress energy tensor can be obtained
  if we take e.g.  $q_\mu=(0,0,0,q)$:
\bea
T^0_{\,\,\,0}&=&\big[\dot\varphi^2/2+ V(\varphi)\big]
 +  \big[(\dot q+q H)^2 +2 q^2 f(\varphi)\big]/2;
 \nonumber\\[5pt]
 - T^k_{\,\,\,k}&=&\big[\dot\varphi^2/2 - V(\varphi)\big]
  \pm 
 \big[(\dot q+q H)^2 - 2 q^2 f(\varphi)\big]/2;
 \label{fT}
 \eea
 % \medskip\noindent
 with  the  contribution of $\w_\mu$ having opposite signs
 for $k=1,2$ (+) and for  $k=3$ (-) and $T^i_j=0$, $i\not=j$.
 The contribution of $q(t)\sim 1/a(t)^{3/2}$ (\ref{qqq}) is
 suppressed by $a(t)^3$  relative to that of $\varphi$.} $\varphi$.

Consider now the equations for perturbations, with
$w_\mu(\vec x, t)=\w_\mu(t)+\delta \w_\mu(\vec x,t)$.
Then eqs.(\ref{oui}) for  $\mu=i$ and $\mu=0$ give
%\medskip
\bea
\ddot\w_i+ H\,\dot\w_i -\frac{1}{a(t)^2} 
\big[\partial_j\partial_j \w_i- \partial_i \partial_j\w_j\big]
+f(\varphi)\, \w_i &=& \partial_i\dot\w_0
+H \partial_i \w_0
\label{K}
\\
\partial_i\dot\w_i -\partial_i \partial_i\w_0 +a(t)^2 \,f(\varphi) \w_0&=&0
\label{I}
\eea

\medskip\noindent
with $\w_\mu=\w_\mu\xt$ and $\varphi=\varphi\xt$.
By applying  $\partial_\mu$ on (\ref{oui}) we find
\bea\label{J}i
\partial_i \dot\w_0 -\frac{1}{a(t)^2} \partial_i\partial_j \w_j+ 3 H \partial_i \w_0+\partial_i D=0
\eea

\medskip\noindent
where  $D\!=\!D\xt$ and $D\!=\! \w_0\partial_0\ln f(\varphi)-1/a(t)^2 \w_j \partial_j \ln f(\varphi)$.
Adding (\ref{K}), (\ref{J}) then
\medskip
\bea
\ddot\w_i + H \dot \w_i -\frac{1}{a(t)^2} \partial_j\partial_j \w_i 
+ f(\varphi) \w_i = -2 H \partial_i \w_0 -\partial_i D.
\eea
This gives for perturbations $\delta\w_\mu$ a linear differential equation:
\be\label{p3}
\delta\ddot w_i + H \delta\dot\w_i -\frac{1}{a(t)^2} \partial_j\partial_j \delta\w_i +
f(\varphi(t)) \delta\w_i=
-(2 H+\partial_0\ln f(\varphi(t))) \partial_i \delta\w_0
- f^\prime(\varphi(t)) \w_j(t) \sigma_{ij}(t) 
\ee
Here $\sigma_{ij}\xt=
[\delta_{ij}+ 1/(a(t)^2 f(\varphi(t)))\,\, \partial_i\partial_j]\, \delta\varphi\xt$;
notice that in general case of $\w_i(t)\not\!=\!0$ (\ref{p3}) is 
``mixing'' $\delta\varphi$ and $\delta\w_\mu$. However, this mixing is absent in our
FRW case of $\w_\mu(t)\!=\!0$.

Further, from  remaining (\ref{I})
\medskip
\bea
\partial_i\, \delta\dot\w_i(\vec x,t)-\partial_i\partial_i\,\delta\w_0(\vec x,t) +a(t)^2\,
 f(\varphi(t))\,\delta\w_0(\vec x,t)=0.
 \eea
or, in Fourier modes
\bea
\delta\w_0(\vec k, t)=-i\frac{k_j \delta\dot\w_j (\vec k, t)}{k^2+a^2 f(\varphi(t))}.
\eea
We separate the  perturbations into parallel and orthogonal directions to  $\vec k$ (taken along OZ):
\bea
\delta \vec\w^\parallel=\frac{\vec k (k_i \delta \w_i)}{k^2}, \qquad
\delta\vec\w^\perp=\delta \vec\w -\delta\vec\w^\parallel.
\eea
We introduce the physical perturbations $\delta q_\mu\xt=(1/a(t))\,\delta\w_\mu\xt$ and  express
(\ref{p3}) in terms of the Fourier modes $\delta q_\mu(\vec k,t)$ defined by
$\delta q_\mu\xt =\int d^3k/(2\pi)^{3/2} \delta q_\mu(k,t) \exp(i\vec k\vec x)$. We find
for the Fourier modes of parallel $\delta q_z^\parallel(\vec k,t)$
and orthogonal $\delta q^\perp_j(\vec k,t)$ directions\footnote{
  The equations for $\delta \w^\parallel$ ($\delta \w^\perp$) are similar
  to those for $\delta q^\parallel$ ($\delta q^\perp$)
   but with coefficient $3 H$ replaced by $H$ and
  without any $H$-dependence inside the brackets multiplying $\delta q^\parallel$ ($\delta q^\perp$)
  respectively.} 
 \bea
\label{a1}
\delta\ddot q^{\parallel}_z+\delta\dot q_z^\parallel\, \Big[
3H+ \theta_1\Big]+\Big[\,\frac{k^2}{a^2}+2 H^2+ H \theta_1+
f(\varphi(t)) \Big]\delta q_z^\parallel&=&\theta_2,
\\
\delta\ddot q_j^\perp+ 3 H\,\delta\dot  q_j^\perp+\Big[\,\frac{k^2}{a^2}+2H^2+
f(\varphi(t))\,\Big]\delta q_j^\perp&=&0, \quad j=1,2.
\label{a2}
\eea
where
\bea
\theta_1=\frac{k^2 \big( 2 H+\partial_0 \ln f(\varphi(t)))}{k^2+a^2 f(\varphi(t)\big)},
\qquad
\theta_2=-\delta\varphi_k\, q_z(t) f^\prime(\varphi(t)) \Big(1- \frac{k^2}{a^2 f(\varphi(t))}\Big).
\eea
Eqs.(\ref{a1}), (\ref{a2}) are similar to those in  \cite{Di1} (eqs.21, 22, 67)
 except  an extra  $\varphi$-dependent correction to the mass ($\sim M^2$) of
 $\w_\mu$ that induces $\theta_2$ and a time derivative in $\theta_1$.

 Eq.(\ref{a2}) is similar to that for the scalar field perturbations, eq.(\ref{j3}).
 We expect perturbations $\delta q_j^\perp$ be generated 
 if their  effective mass $m^2=2 H^2+f(\varphi(t))<H^2$. This condition
 is not respected  since $f(\varphi)>0$.
 The power spectrum is  exponentially suppressed, as for the
scalar field,  eq.(\ref{j3}) with $\cV''\!\ra\! f(\varphi(t))+2H^2$
with  $\nu$  imaginary and eq.(\ref{zx3}).

Similar considerations apply to the perturbations to the parallel 
(longitudinal) mode  of $\w_\mu$.  
For our FRW-compatible background $q_z^\parallel(t)=0$ ($\w_\mu(t)=0$), hence
$\theta_2=0$. Therefore, there is no mixing of $\delta q_z^\parallel$ and $\delta\varphi_k$ perturbations
in (\ref{a1}), in agreement with the earlier similar finding, see discussion around eq.(\ref{j3prime}).
Note also that if  $k^2\ll a^2 f(\varphi)$, $\delta q_z^\parallel$
has an equation  similar to the transverse modes,
with $\theta_1\sim 0$ (with $H\sim$constant, $\dot\varphi^2\sim -2 \dot H^2 M^2$).
Similar to the transverse case,
the effective mass $m^2=2H^2+ H \theta_1 +f(\varphi(t))$
of  $\delta q_z^\parallel$ is again larger  than $H$ and its
generation is exponentially suppressed. We see again that in the FRW case one can
ignore the effect of $\delta\w_\mu$ and of coupling of $\w_\mu-\varphi$ on  $\delta\phi_k$.

In a general  background case
$q_z^\parallel(t)\!\not=\!0$, then $\theta_2\!\not=\!0$; then a mixing
of perturbations of $\varphi$ and of longitudinal mode of $\w_\mu$ exists in (\ref{a1})
due to  coupling $f(\varphi) w_\mu w^\mu$, eq.(\ref{gfi}).
However, even in this case,   $q_z^\parallel$   is suppressed by the scale factor,
due to  eq.(\ref{qqq}), and thus the same is true for the mixing.

\vspace{0.5cm}
\noindent
{\bf Acknowledgements:}
This work was partially supported 
by a grant from the Romanian Ministry of Education and Research,  % CNCS - UEFISCDI,
project number PN-III-P4-ID-PCE-2020-2255. %, within PNCDI III.

\end{document}